\documentclass[10pt,conference]{IEEEtran}
\usepackage{cite}
\usepackage{amsmath,amssymb,amsfonts}
\usepackage{algorithmic}
\usepackage{graphicx}
\usepackage{textcomp}
\usepackage{xcolor}
\usepackage[hyphens]{url}
\usepackage{fancyhdr}
\usepackage{hyperref}
\usepackage{subcaption}  % 서브피규어용
\usepackage{tikz}
\usepackage{verbatim}
\usepackage{multirow}
\usepackage{tabularx}
\usepackage[most]{tcolorbox}  % 필요한 기능 다 포함
\usepackage{siunitx}

\newcommand*\ecircled[1]{\tikz[baseline=(char.base)]{
            \node[shape=circle,draw,inner sep=0.5pt] (char) {#1};}}

\newcommand*\ecircledblack[1]{\tikz[baseline=(char.base)]{
    \node[shape=circle, draw, fill=black, text=white, inner sep=0.5pt] (char) {#1};}}

\newtcolorbox{graybox}[1][]{
    colback=gray!20,      % 배경색 (회색 20%)
    colframe=gray!50,     % 테두리색 (회색 50%)
    boxrule=0.5pt,        % 테두리 두께
    arc=2mm,              % 모서리 둥글기
    left=5pt, right=5pt, top=3pt, bottom=3pt,
    #1
}

\newcommand{\hpcayear}{2027}

\newcommand{\hpcasubmissionnumber}{545}
\title{HyPAIC: Personal AI Computer with High Bandwidth Flash for High Performance Trillion-Parameter LLM Inference
}%%%%%%%%%%%%%%%%%%%%%%%%%%%%%%%%%%%%%%%%

\def\hpcacameraready{} % Unto build camera-ready version

\newcommand\hpcaauthors{Minho Ha, Myunghyun Rhee, Euiseok Kim, and Hoshik Kim}
\newcommand\hpcaaffiliation{SK hynix Inc.}
\newcommand\hpcaemail{\{minho1.ha, myunghyun.rhee, euiseok.kim, hoshik.kim\}@sk.com}

\author{
  \ifdefined\hpcacameraready
    \IEEEauthorblockN{\hpcaauthors{}}
      \IEEEauthorblockA{
        \hpcaaffiliation{} \\
        \hpcaemail{}
      }
  \else
    \IEEEauthorblockN{\normalsize{HPCA \hpcayear{} \textcolor{blue}{Revision}
      \textbf{\#\hpcasubmissionnumber{}}} \\
      \IEEEauthorblockA{
        Confidential Draft \\
        Do NOT Distribute!!
      }
    }
  \fi 
}

\begin{document}
\maketitle

%Enables the camera ready header and footer
\ifdefined\hpcacameraready 
  \thispagestyle{empty}
  \pagestyle{empty}
\else
  \thispagestyle{plain}
  \pagestyle{plain}
\fi

\newcommand{\hpcaheight}{0mm}
\ifdefined\eaopen
\renewcommand{\hpcaheight}{12mm}
\fi

%%%%%%%%%%%%%%%%%%%%%%%%%%%%%%%%%%%%%%%%
%%%%%%%% -- PAPER CONTENT STARTS -- %%%%%%%%%

\begin{abstract}
Large language models (LLMs) have undergone rapid adoption, expanding from simple chatbots and reasoning tasks to agentic AI, where models directly utilize tools to solve complex problems. 
To achieve better problem-solving capabilities, most modern frontier LLMs adopt a mixture-of-experts (MoE) architecture, pushing model sizes into the trillion-parameter regime. 
While cloud-based services have been predominant for large-scale LLM inference, there is increasing demand for on-premises environments due to cost and security issues. 
In response, GPU manufacturers have introduced personal AI computers (PAICs). 
However, existing PAICs equipped with LPDDR-based unified main memory suffer from severe capacity limitations, restricting the size of LLMs that can be hosted on a single device. 
If most of the memory is used to store model weights, less capacity remains for KV cache storage, reducing the supported context length.

In this paper, we propose HyPAIC (Hyper-PAIC), a PAIC architecture that incorporates high bandwidth flash (HBF) into PAICs to overcome these memory constraints, introducing a method to store read-only model weights in HBF and mutable runtime data in LPDDR.
HBF provides over 10$\times$ the capacity of high bandwidth memory (HBM) alongside comparable TB/s-scale bandwidth. 
% Crucially, its inherent drawbacks—namely, high access latency, low write endurance, high power consumption, and page-level granularity—are naturally mitigated when integrated into PAICs, which are inherently designed for low batch, single-user workloads.
HyPAIC addresses HBF’s access latency, limited write endurance, and page-level access granularity through prefetching, read-only weight placement, and coarse-grained weight accesses, respectively. 
Its power and cooling requirements remain significant implementation challenges.
Through simulation-based evaluation, we demonstrate that \textcolor{black}{a single} HyPAIC improves end-to-end (E2E) system throughput by up to \textcolor{black}{1.04}$\times$ and TDP-normalized throughput by up to \textcolor{black}{3.59}$\times$ during 1T MoE model inference\textcolor{black}{, compared with the aggregate performance of eight baseline systems without HBF.
In other words, a single HyPAIC approaches the raw performance of eight baseline systems while offering far superior power efficiency.}
For small model inference, HyPAIC increases the maximum number of concurrent tasks that maintain at least 50 decode tokens per second per task by up to \textcolor{black}{4}$\times$, significantly enhancing both user experience and productivity.
Although HBF specifications are still evolving, this study demonstrates that integrating HBF into PAICs provides strong architectural synergy.
\end{abstract}

\section{Introduction}
\label{sec:intro}

Large language models (LLMs) have quickly been adopted, growing from simple chatbots and reasoning tasks into agentic AI systems that directly leverage tools to solve complex problems~\cite{deepseek-v4,qwen3.5,team2026kimi}.
Most frontier LLMs today adopt mixture-of-experts (MoE) architectures to enhance problem-solving capabilities, pushing model sizes into the trillions of parameters.
The latest MoE-based open-source models, DeepSeek-V4-Pro~\cite{deepseek-v4} and Kimi~K2.5~\cite{team2026kimi}, are reported to have 1.6 trillion (T) and 1T parameters, respectively, while the closed-source models OpenAI GPT-5.5 Pro and Claude Fable 5 are estimated to have up to 5.3T and 3.5T parameters, respectively~\cite{li2026incompressible}.
These models are known to demonstrate performance comparable to or exceeding human performance on several benchmarks.

To date, research on LLM serving systems in data centers has primarily focused on two metrics: 
% system throughput (the number of tokens generated concurrently), which dictates operating costs
aggregate system throughput (the total number of output tokens generated per second across all requests), which influences serving cost per token~\cite{kwon2023efficient,agrawal2023sarathi}; and user experience (how quickly individual tasks are processed)~\cite{lie2024wafer,abts2022groq}.
Since the majority of users perform large-scale LLM inference through cloud-based services, such as ChatGPT~\cite{chatgpt}, Gemini~\cite{gemini}, and Claude~\cite{claude}, these research priorities have dominated the field; however, recently, there has been an increasing demand for on-premises environments due to cost and security issues.

Recently, in response to this demand, GPU manufacturers are competitively releasing personal AI computers (PAICs).
NVIDIA unveiled DGX Spark~\cite{dgx-spark} and RTX Spark~\cite{rtx-spark}, which are based on GB10~\cite{skende202617}, and AMD released Ryzen AI Halo~\cite{ryzen-ai-halo}.\footnote{We exclude server-scale computers like the DGX Station~\cite{dgx-station}, which is equipped with a GB300, demonstrating impressive performance with 10 PFLOPS and 748 GB of memory capacity. 
However, while PAICs can be purchased for a few thousand dollars, DGX Station is known to have a minimum purchase price of around \$80,000~\cite{dgx-station-price}, so we believe it is difficult for an individual to buy these server-scale computers.}
These PAICs are equipped with an SoC in which the CPU and GPU are connected via chiplets, and use LPDDR as the unified main memory.
Due to the capacity limitations of LPDDR, the size of the LLMs that can be loaded onto a single PAIC is limited. 
DGX Spark's 128 GB of LPDDR, for example, can store up to \textcolor{black}{227} billion (B) \textcolor{black}{NVFP4} parameters, without accounting for runtime data. 
Since LPDDR must store not only model weights but also key-value (KV) caches and activation values, larger models result in smaller available KV cache space. 
This reduces the context length that the model can support.
% There are ways to run larger models and store longer contexts. 
Several methods exist to run larger models and store longer contexts.
One approach is to utilize the built-in SSD, and another is to scale out by connecting multiple PAICs.
% However, utilizing SSD presents the problem of rapidly dropping performance due to its low bandwidth, and using multiple PAICs leads to increased costs and power consumption.
However, utilizing an SSD significantly degrades performance due to its low bandwidth, while scaling to multiple PAICs incurs higher costs and power consumption.
Therefore, a method that is both effective and efficient is required to simultaneously expand the memory capacity and bandwidth of PAICs.

One promising method is to utilize high bandwidth flash (HBF).
HBF is a device originally proposed to enable data center accelerators to effectively process larger volumes of data~\cite{ilkbahar2026high}.
Similar to high bandwidth memory (HBM), which implements high bandwidth by stacking DRAM dies on a logic die and using through silicon vias (TSVs), HBF supports high bandwidth by stacking NAND flash dies on a logic die and using TSVs.
The advantages of HBF include \ecircled{1} a capacity more than 10$\times$ larger than that of HBM, and \ecircled{2} a bandwidth on a TB/s scale comparable to that of HBM. 
Its disadvantages include \ecircledblack{1} higher access latency compared to HBM, \ecircledblack{2} low write endurance, \ecircledblack{3} higher power consumption than HBM, and \ecircledblack{4} page-level granularity~\cite{ha2026h}.
% Since HBF is a device with distinct advantages and disadvantages, to utilize it effectively, the characteristics of the workload should be thoroughly understood to maximize its strengths and minimize its weaknesses.
% Considering the advantages and disadvantages of HBF, we determine that a workload for inferring large-scale MoE models with a low batch size is suitable for HBF.
% However, running large-scale MoE inference with a low batch size is not suitable for data centers, where system throughput must be boosted by maximizing batch size.
% In other words, data centers are not suitable for performing large-scale MoE inference with a low batch size to suit HBF.
Given these distinct advantages and disadvantages, workloads mapped to HBF should satisfy four conditions: 1) large capacity and high bandwidth requirements, 2) read-heavy access pattern, 3) coarse-grained access, and 4) sparse access.
Considering these conditions, HBF is particularly well suited to large-scale MoE inference at low batch sizes. 
However, this operating regime is less aligned with conventional throughput-oriented data center deployments, which rely on maximizing batch sizes to boost aggregate system throughput. 
On the other hand, integrating HBF into a PAIC, which is inherently designed for low batch, single-user workloads, presents an ideal architectural synergy. 
This integration overcomes the memory constraints of existing PAICs while leveraging the advantages of HBF and mitigating its disadvantages.

% By exploiting this hardware-workload alignment, we propose a novel PAIC architecture incorporating HBF for large-scale MoE inference. 
By exploiting this hardware-workload alignment, we propose a novel PAIC architecture, \textbf{HyPAIC (Hyper-PAIC)}, incorporating HBF for large-scale MoE inference. 
Considering the current hardware limitations of PAICs and the characteristics of HBF-suitable applications, performing large-scale MoE inference by integrating HBF into the PAIC can overcome PAIC's limitations while fully exploiting HBF's strengths.
From the PAIC perspective, HBF increases both memory capacity and bandwidth, thereby improving throughput and TDP-normalized throughput.
From the HBF perspective, low batch MoE inference satisfies all four workload requirements identified above.
% Consequently, HBF can be used most effectively.
% To the best of our knowledge, this paper is the first instance of applying HBF to PAIC.

The key contributions of this paper are as follows:

\begin{itemize}
    \item We quantitatively identify architectural problems that occur when performing LLM inference using the existing PAIC (Sec.~\ref{subsec:paic_limitations}).
    \item We systematically classify workload characteristics suitable for HBF (Sec.~\ref{subsec:hbf_workload}).
    \item We propose a new PAIC architecture, HyPAIC, equipped with HBF and present a method for utilizing this architecture (Sec.~\ref{sec:paic_hbf}).
    % \item We develop an architecture simulator that reflects both the architectural nature of PAIC and the characteristics of the latest MoE models (Sec.~\ref{sec:paic_sim}).
    \item Through simulation-based evaluation, we show that HyPAIC can improve end-to-end (E2E) system throughput by up to \textcolor{black}{1.04}$\times$ and E2E TDP-normalized throughput by up to \textcolor{black}{3.59}$\times$ during 1T MoE model inference. 
    Additionally, we show that the maximum task concurrency while maintaining at least 50 decode tokens per second (TPS) per task, increases by up to \textcolor{black}{4}$\times$ for 35B MoE model inference.
    \textcolor{black}{We also analyze performance sensitivity to HBF bandwidth and prefetch hit rate.}
    \textcolor{black}{Finally, we demonstrate the superiority of the proposed HyPAIC through comparisons with other HBF hybrid architectures and iso-bandwidth experiments (Sec.~\ref{sec:eval}).}
\end{itemize}

\section{Background}
\label{sec:background}

\subsection{State-of-the-Art LLM Trends}
\label{subsec:sota_llm_trends}

\begin{figure}[t]
  \centering
  \includegraphics[width=1.0\linewidth]{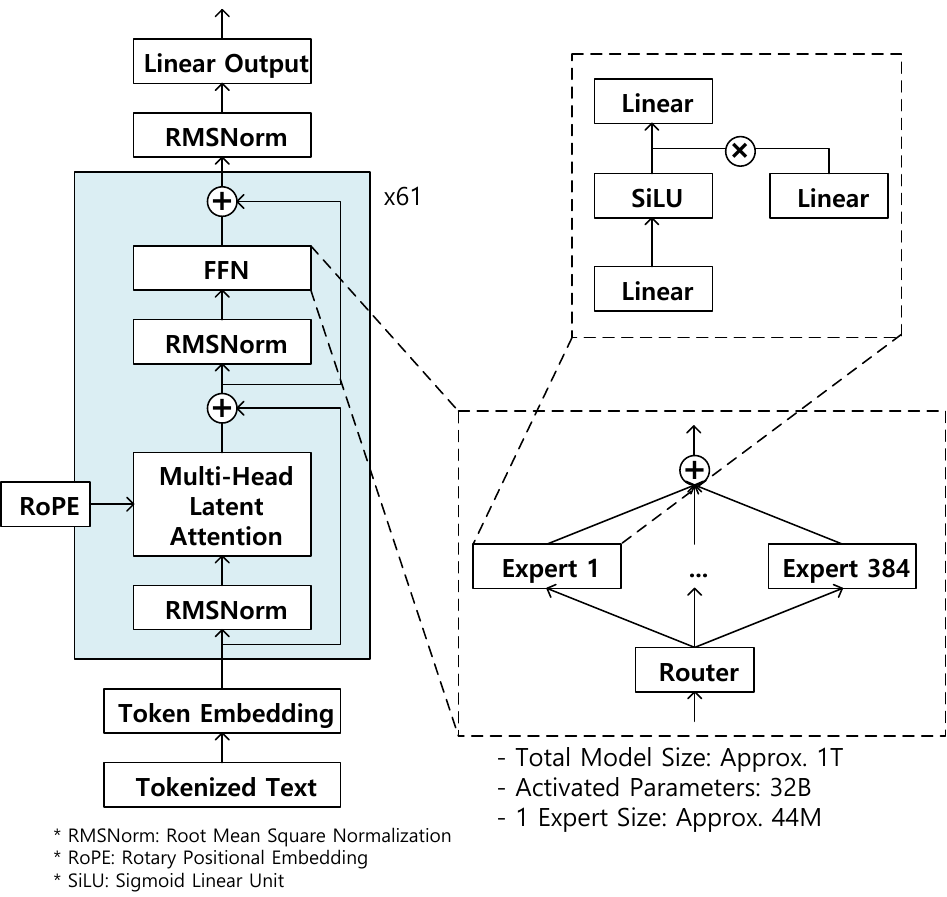} 
  \caption{MoE architecture example (Kimi~K2.5~\cite{team2026kimi}).}
  \vspace{-1pc}
  \label{fig:moe}
\end{figure}

High-performance frontier LLMs are scaling toward trillions of parameters (Table~\ref{tab:llm}).
A widely used approach for scaling model capacity is the MoE technique~\cite{fedus2022switch}.
The MoE technique employs multiple experts in each feedforward network (FFN) layer, and during the actual inference process, only the top-k most relevant experts are selected for processing. 
For example, in the case of Kimi~K2.5~\cite{team2026kimi}, the total parameter count is 1T, and only 32B parameters are activated (Fig.~\ref{fig:moe}).
This allows the model size to be flexibly scaled while keeping computations low.
Even when using the \textcolor{black}{NVFP4} data type, DeepSeek-V4-Pro~\cite{deepseek-v4} and Kimi~K2.5~\cite{team2026kimi} require 900~GB and 562.5~GB, respectively, solely for storing model weights. 
%(Fig.~\ref{fig:weight_plot}).

\begin{table}[t]
\caption{Intelligence scores based on LLM model architecture type and number of parameters}
\vspace{-0.5pc}
\begin{center}
\begin{tabular}{|c|c|c|c|}
\hline
\textbf{Models} & \textbf{\begin{tabular}[c]{@{}c@{}}Total \\ Parameters\end{tabular}} & \textbf{\begin{tabular}[c]{@{}c@{}}Dense\\ /MoE\end{tabular}} & \textbf{\begin{tabular}[c]{@{}c@{}}MMLU Pro\\ Score\end{tabular}} \\ \hline
DeepSeek-V4-Pro~\cite{deepseek-v4} & 1.6T                                                                & MoE                                                           & 87.5\%           \\ \hline
Kimi~K2.5~\cite{team2026kimi}       & 1T                                                                  & MoE                                                           & 87.1\%           \\ \hline
Llama 3.1 405B~\cite{llama3.1}  & 405 B                                                                & Dense                                                         & 73.3\%           \\ \hline
Llama 3.1 70B~\cite{llama3.1}    & 70 B                                                                   & Dense                                                         & 66.4\%           \\ \hline
\end{tabular}
\end{center}
\label{tab:llm}
\vspace{-1pc}
\end{table}

\begin{comment}
\begin{figure}[t]
  \centering
  \includegraphics[width=0.9\linewidth]{./figure/weight_plot_v0.1.pdf} 
  \caption{Weight sizes of the latest open-source LLMs.}
  \label{fig:weight_plot}
\end{figure}
\end{comment}

\subsection{LLM Inference}
\label{subsec:llm_inf}

LLM inference is divided into two phases: a prefill phase, which receives input tokens, understands their content, and generates the first token; and a decode phase, which subsequently generates tokens one by one in an autoregressive manner~\cite{nvidia_llm_inference}. 
% (Fig.~\ref{fig:llm_inference}).
Traditionally, the prefill phase is known to be compute-bound because it performs general matrix-matrix multiplication (GEMM) operations on the entire sequence of input tokens, while the decode phase is memory-bound because it performs general matrix-vector multiplication (GEMV) operations only on new tokens.
% The key factor that makes the decode phase memory-bound is the KV cache.
% Because the KV cache generated by the operation is continuously stored and reused, the amount of computation in the decode phase is drastically reduced.
The key factor that renders the decode phase memory-bound is the KV cache.
Since the KV cache generated during the prefill phase is continuously stored and reused, the amount of computation required for subsequent decode steps is significantly reduced.

Due to these distinct computational characteristics of prefill and decode during the LLM inference process, prefill-decode (PD) disaggregation, which separates nodes handling prefill from nodes handling decode, is common in data centers composed of multiple accelerator clusters~\cite{patel2024splitwise,zhong2024distserve,qin2025mooncake}.
However, in resource-constrained environments like PAICs~\cite{dgx-spark,rtx-spark,ryzen-ai-halo}, PD co-location is often necessary.

\begin{comment}
\begin{figure}[t]
  \centering
  \includegraphics[width=1.0\linewidth]{./figure/llm_inference_v0.1.pdf} 
  \caption{Example of LLM inference.}
  \label{fig:llm_inference}
  \vspace{-1pc}  
\end{figure}
\end{comment}

\subsection{Personal AI Computer}
\label{subsec:PAIC}

\begin{table}[t]
\caption{Specifications of released PAICs}
\vspace{-0.5pc}
\begin{center}
\begin{tabular}{|c|c|c|}
\hline
                 & \textbf{NVIDIA DGX Spark~\cite{dgx-spark}}                                                                    & \textbf{AMD Ryzen AI Halo~\cite{ryzen-ai-halo}}                                                       \\ \hline
\textbf{CPU}     & \begin{tabular}[c]{@{}c@{}}20-core Arm\\ (10 Cortex-X925 \\ + 10 Cortex-A725)\end{tabular} & 16-Core  Zen 5 Architecture                                                      \\ \hline
\textbf{GPU}     & \begin{tabular}[c]{@{}c@{}}Blackwell Architecture\\ 500 TOPS (NVFP4, Dense)\end{tabular}     & \begin{tabular}[c]{@{}c@{}}RDNA 3.5 Architecture\\ 60 TFLOPS (FP16)\end{tabular} \\ \hline
\textbf{NPU}     & -                                                                                            & \begin{tabular}[c]{@{}c@{}}XDNA 2 Architecture\\ 50 TOPS\end{tabular}            \\ \hline
\textbf{Memory}  & \begin{tabular}[c]{@{}c@{}}LPDDR5x 8533 MT/s\\ 128 GB \& 273 GB/s\end{tabular}               & \begin{tabular}[c]{@{}c@{}}LPDDR5x 8000 MT/s\\ 128 GB \& 256 GB/s\end{tabular}   \\ \hline
\textbf{Storage} & Up to 4 TB                                                                                   & Up to 2 TB                                                                       \\ \hline
\textbf{Network} & 200 GbE                                                                                       & 10 GbE                                                                           \\ \hline
\textbf{TDP} & 240 W                                                                                       & 120 W                                                                       \\ \hline
\end{tabular}
\end{center}
\label{tab:paics}
\vspace{-1pc}
\end{table}

\begin{figure}[t]
  \centering
  \includegraphics[width=0.9\linewidth]{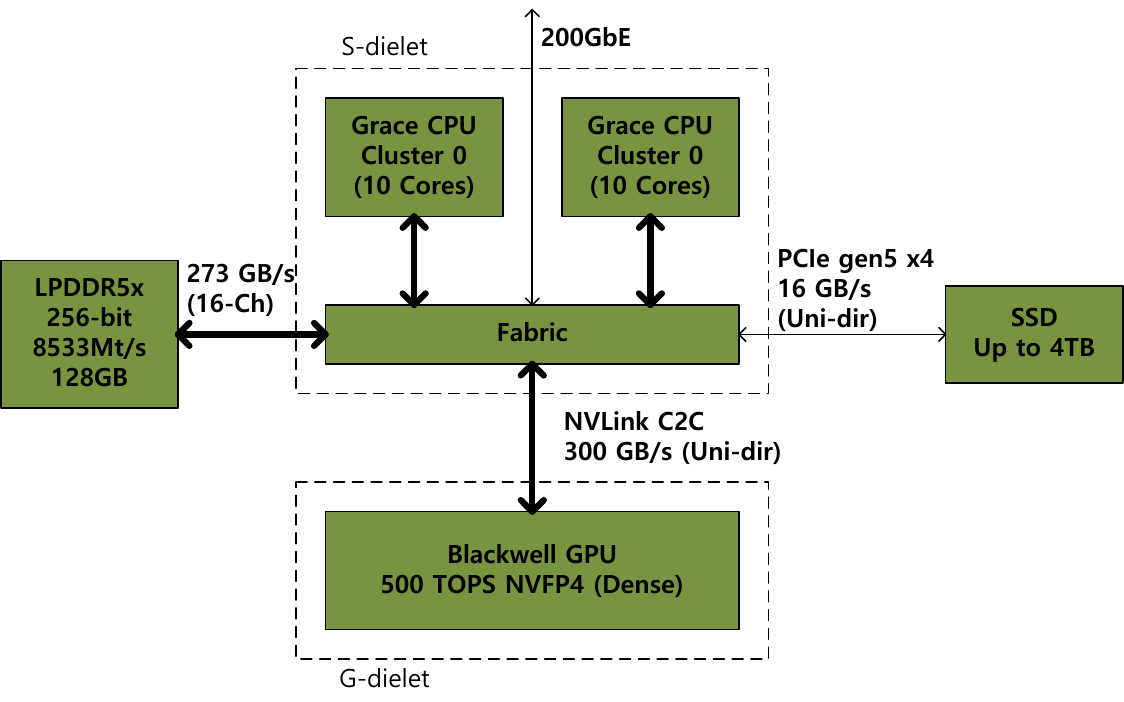} 
  \caption{System architecture of DGX Spark~\cite{dgx-spark}.}
  \label{fig:paic_baseline}
\end{figure}

It is common to use frontier LLMs running in data centers via cloud-based services~\cite{chatgpt,gemini,claude}.
As an alternative to this method, NVIDIA and AMD have released their respective PAICs, the DGX Spark~\cite{dgx-spark} and Ryzen AI Halo~\cite{ryzen-ai-halo}, to enable high-performance LLM operations in local environments (Table~\ref{tab:paics}, Fig.~\ref{fig:paic_baseline}).
The target consumers for PAICs include engineers and researchers who wish to fine-tune AI models, deploy AI agents, and serve LLMs in local environments.

Several factors have driven the emergence of PAICs.
First, they reduce the exorbitant costs of cloud services.
Previously, running large-scale LLMs mandated the use of cloud-based services.
The cost is substantial; for example, the token generation cost for Claude's latest model, Fable 5, is up to \$50 per million (M) output tokens~\cite{claude}. 
This imposes a significant financial burden on users.
Therefore, PAICs can be an attractive solution as they allow for lower token generation costs after the initial equipment purchase cost has been paid.
For example, since the selling price of DGX Spark is \$4,699~\cite{dgx-spark-price}, using it for three years would require \$131 per month. 
Assuming the average household electricity rate in the U.S. is \$0.188/kWh (as of April 2026)~\cite{elect_power_month}, the monthly cost including 24-hour electricity usage fees would be \$163 per month.
If 5M tokens (assuming an input-to-output ratio of 4:1 and an 80\% prompt cache hit) are used per day using Fable 5 (priced at \$10/\$50/\$1 for 1M input tokens/output tokens/prompt cache respectively)~\cite{claude}, \$1,836 is required per month.
For users who frequently run LLM inference, it would be much more economically advantageous to build a system using a PAIC.
A second motivation is data security.
Sending proprietary or sensitive corporate data to an external cloud can be a security risk.
This issue can be resolved by running the LLM in an on-premises environment isolated from the external network.

\subsection{High Bandwidth Flash}
\label{subsec:hbf}

\begin{figure}[t]
  \centering
  \includegraphics[width=0.8\linewidth]{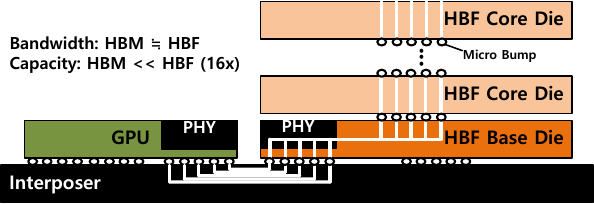} 
  \caption{Concept of HBF.}
  \vspace{-1pc}
  \label{fig:hbf}
\end{figure}

Modern AI data center accelerators have been equipped with HBM to achieve high throughput~\cite{gb200,verarubin,mi355x,mi455x,tpuv8,mtia}.
Due to the demand for larger memory capacity in these accelerators, HBF has been proposed~\cite{ilkbahar2026high}.
HBF essentially has an architecture similar to that of HBM (Fig.~\ref{fig:hbf}).
NAND flash dies are stacked on top of logic dies, and data is transmitted at high bandwidth through TSVs penetrating the stacked dies.
The accelerator and HBF cubes are integrated into a 2.5D configuration via an interposer.

HBF has clear advantages and disadvantages~\cite{ha2026h}.
HBF has two advantages: \ecircled{1} a capacity more than 10$\times$ larger than that of HBM and \ecircled{2} the provision of HBM-level bandwidth.
In this paper, we assume 1~TB/s per HBF cube with 576 GB, which is 16$\times$ larger capacity than HBM3e 12-Hi (actual product specifications may vary).
On the other hand, HBF also has several disadvantages due to the inherent limitations of NAND flash~\cite{ha2026h}: \ecircledblack{1} high read/write latency, \ecircledblack{2} low write endurance, which is critical to device lifespan, \ecircledblack{3} high read/write power consumption, and \ecircledblack{4} page-level granularity, which can cause read/write amplification.
Therefore, HBF must be used with appropriate workloads in a suitable environment by carefully considering its benefits and drawbacks.
If used in inappropriate environments for unsuitable workloads, it can lead to performance degradation and significantly shorten device lifetime.

\section{Motivation}
\label{sec:motivation}

\subsection{Limitations of Personal AI Computer}
\label{subsec:paic_limitations}

Existing PAICs are equipped with unified LPDDR5x-based main memory~\cite{dgx-spark,rtx-spark,ryzen-ai-halo} (Table~\ref{tab:paics}).
This capacity is significantly insufficient to run trillion-parameter models like Kimi~K2.5~\cite{team2026kimi}.
For example, assuming the 128 GB supported by DGX Spark and AMD Ryzen AI Halo, models with at most \textcolor{black}{227B} parameters can be hosted, even for \textcolor{black}{NVFP4} models.
Since the KV cache and activation values should be stored in the remaining capacity after the models are loaded, LPDDR5x also has issues maintaining long task contexts.
To load larger models and store longer contexts, users can \ecircled{1} use SSDs or \ecircled{2} use multiple PAICs.
However, each presents problems: \ecircled{1} performance degradation due to the limited bandwidth of SSDs, and \ecircled{2} increased cost and power consumption.

\begin{figure*}[t]  % [t] = 페이지 상단, [b] = 하단, [p] = 단독 페이지
    \centering
    
    % 첫 번째 행
    \begin{subfigure}[b]{0.32\textwidth}
        \centering
        \includegraphics[width=\linewidth]{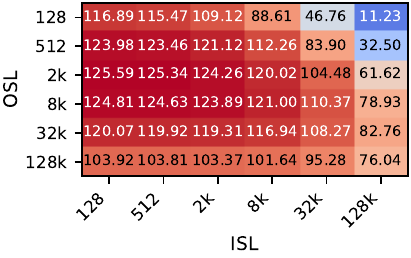}
        \caption{Qwen3.5-35B-A3B (1 DGX Spark)}
        \label{fig:baseline_perf_sub1}
    \end{subfigure}
    \hfill  % 열 간격 균등 분배
    \begin{subfigure}[b]{0.32\textwidth}
        \centering
        \includegraphics[width=\linewidth]{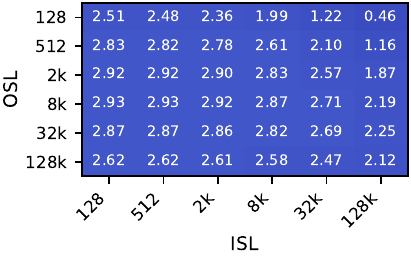}
        \caption{Qwen3.5-397B-A17B (1 DGX Spark)}
        \label{fig:baseline_perf_sub2}
    \end{subfigure}
    \hfill
    \begin{subfigure}[b]{0.32\textwidth}
        \centering
        \includegraphics[width=\linewidth]{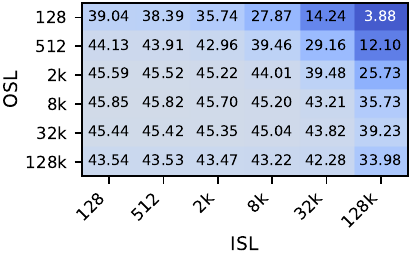}
        \caption{Qwen3.5-397B-A17B (2 DGX Sparks)}
        \label{fig:baseline_perf_sub3}
    \end{subfigure}

    \vspace{0.5pc}
    
    % 두 번째 행
    \begin{subfigure}[b]{0.32\textwidth}
        \centering
        \includegraphics[width=\linewidth]{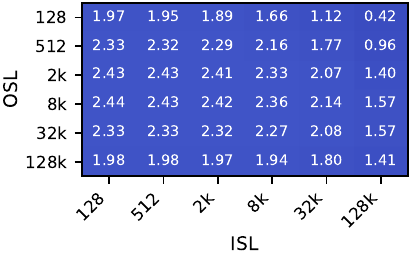}
        \caption{Kimi~K2.5 (\textcolor{black}{2 DGX Sparks})}
        \label{fig:baseline_perf_sub4}
    \end{subfigure}
    \hfill
    \begin{subfigure}[b]{0.32\textwidth}
        \centering
        \includegraphics[width=\linewidth]{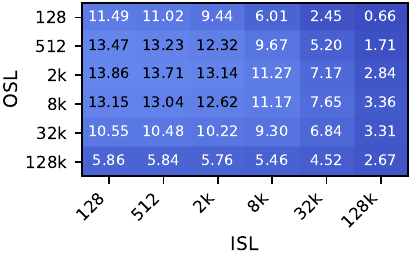}
        \caption{Kimi~K2.5 (\textcolor{black}{4} DGX Sparks)}
        \label{fig:baseline_perf_sub5}
    \end{subfigure}
    \hfill
    \begin{subfigure}[b]{0.32\textwidth}
        \centering
        \includegraphics[width=\linewidth]{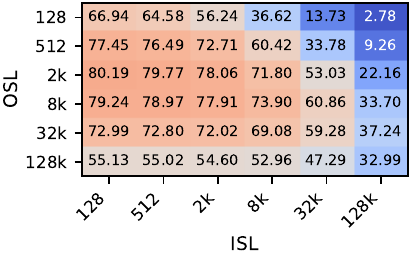}
        \caption{Kimi~K2.5 (\textcolor{black}{8} DGX Sparks)}
        \label{fig:baseline_perf_sub6}
    \end{subfigure}

    \caption{\textcolor{black}{E2E throughput according to the LLM types and number of DGX Sparks (Higher is better. black indicates low TPS, and red indicates high TPS.).}}
    \label{fig:baseline_perf}
\end{figure*}

\noindent \textbf{System Throughput Perspective: }
To identify the limitations of the existing PAIC, we use an architectural simulator (see Sec.~\ref{sec:eval} for details) to estimate E2E system throughput for three LLMs (Qwen3.5-35B-A3B, Qwen3.5-397B-A17B~\cite{qwen3.5}, Kimi~K2.5~\cite{team2026kimi}) by varying the input sequence length (ISL) and output sequence length (OSL) (Fig.~\ref{fig:baseline_perf}). 
Unless stated otherwise, all subsequent experiments use DGX Spark~\cite{dgx-spark} as the baseline PAIC.
The results reveal three key findings.
First, when the model fits entirely within LPDDR5x, the E2E TPS is high in most cases. 
Second, 200GbE allows for effective scale-out. If the model size exceeds the capacity of LPDDR5x, building an environment with multiple PAICs can achieve an acceptable E2E TPS (albeit lower than when fitted to a single device). 
Indeed, NVIDIA recommends using dual DGX Sparks when running the 405B model~\cite{dgx-spark}, and for the 1T-parameter Kimi~K2.5 used in this experiment, \textcolor{black}{8} DGX Sparks are required to achieve an acceptable level of TPS.
Third, when SSD access occurs, performance drops sharply to single-digit TPS, rendering it impractical for real-world deployments.

\begin{graybox}
\textit{\textbf{Motivation 1: If a PAIC's main memory capacity and bandwidth are increased, it will become possible to effectively run even trillion-parameter-scale models on a single computer.}}
\end{graybox}

\begin{figure*}[!t]  
    \centering
    
    \begin{subfigure}[b]{0.32\textwidth}
        \centering
        \includegraphics[width=1.0\linewidth]{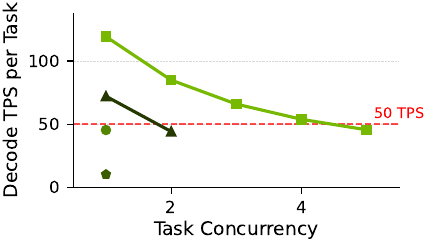}
        \caption{2K/32K}
        \label{fig:tps50_2k32k}
    \end{subfigure}
    \hfill   
    \begin{subfigure}[b]{0.32\textwidth}
        \centering
        \includegraphics[width=1.0\linewidth]{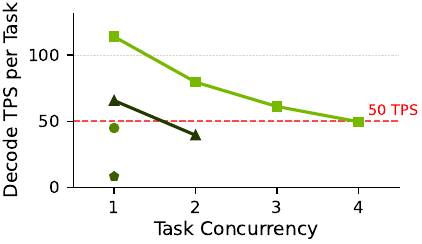}
        \caption{32K/2K}
        \label{fig:tps50_32k2k}
    \end{subfigure}
    \hfill
    \begin{subfigure}[b]{0.32\textwidth}
        \centering
        \includegraphics[width=1.0\linewidth]{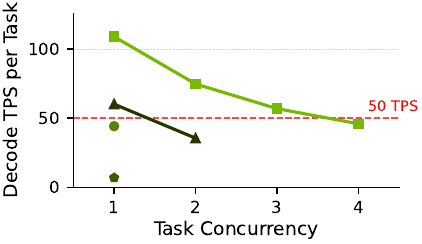}
        \caption{32K/32K}
        \label{fig:tps50_32k32k}
    \end{subfigure}
    
    \vspace{-0.5pc}   
    
    \begin{center}
        \includegraphics[width=0.9\linewidth]{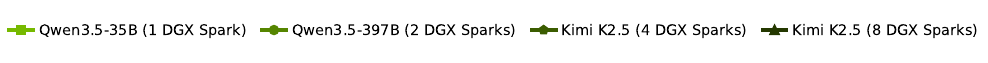}
    \end{center}
    
    \vspace{-1.0pc}   
    
    \caption{\textcolor{black}{Maximum number of concurrent tasks that maintain a decode TPS per task of 50 or higher}.}
    \label{fig:tps50}
\end{figure*}

\noindent \textbf{User Experience Perspective: }
Not only the overall system throughput but also the processing speed of individual tasks needs to be fast.
We estimate the maximum number of concurrent tasks that satisfy a decode TPS per task of 50 or more, which is widely regarded as a smooth interactive experience~\cite{50tps,50tps2}.
Maximum task concurrency is closely related to user productivity.
% As shown in Fig.~\ref{fig:tps50}, when using multiple DGX Sparks, the system can sustain only a single concurrent task to maintain a decode TPS per task of 50 or more. 
\textcolor{black}{When using the existing DGX Spark, the number of tasks that can be processed simultaneously while maintaining a decode throughput of 50 TPS or higher is in the single digits or entirely unfeasible (Fig.~\ref{fig:tps50}).}
Running a larger number of tasks simultaneously reduces the decode TPS per task, which leads to a decline in user experience.
This is caused by insufficient LPDDR5x capacity and bandwidth.

\begin{graybox}
\textit{\textbf{Motivation 2: If PAIC's main memory capacity and bandwidth are increased, it will be possible to process a larger number of tasks simultaneously while maintaining 50 decode TPS per task. Higher task concurrency can improve user productivity.}}
\end{graybox}

\subsection{HBF-suitable Workload}
\label{subsec:hbf_workload}

As summarized in Sec.~\ref{subsec:hbf}, the workloads suitable for HBF are limited due to the clear advantages and disadvantages of HBF.
Considering the strengths and weaknesses of HBF, we summarize the conditions that a workload suitable for HBF must meet as follows.

\noindent \textbf{Large Capacity with High Bandwidth Required: }
The workload should be bound by memory capacity.
More specifically, this condition applies when, unlike using HBM alone which cannot store all the data required for workload processing and necessitates the use of multiple accelerators, utilizing HBF enables data storage using a smaller number of accelerators.
At the same time, since it is memory bandwidth bound, memory bandwidth should affect workload processing performance.

\noindent \textbf{Read-heavy Access Pattern: }
To overcome the low write endurance of HBF, the workload should ideally exhibit a read-heavy access pattern.
According to recent studies, HBF can be used only in situations where write operations are very restricted~\cite{kyung2026high}, and it appears to affect the lifespan of the device when used without restrictions~\cite{son2026exploring}.
The most straightforward approach to mitigate the low write endurance issue is to utilize HBF as a heterogeneous memory and store only read-only data in it~\cite{ha2026h}.

\noindent \textbf{Coarse-grained Access: }
Since HBF is based on NAND flash, the data access granularity is a 4 KB page~\cite{ma2026challenges}.
Therefore, if the minimum access granularity of a workload is smaller than 4 KB, amplification may occur during access.
To prevent this, the granularity of the data logically used within the workload must be 4 KB or larger.
In addition, to maximize HBF bandwidth utilization, 4 KB pages should fill all HBF channels, and for this, at least tens of KB may be required.

\noindent \textbf{Sparse Access: }
Although HBF is known to have bandwidth comparable to HBM, its capacity is more than 10$\times$ greater, resulting in an imbalance between bandwidth and capacity.
For example, assume an HBF configuration that supports 16$\times$ the capacity of HBM3e while having the same bandwidth.
In memory bandwidth-bound cases, such as the decode phase of LLM inference, the ratio of memory bandwidth to capacity determines the TPS.
Reading the entire 36 GB stored in an HBM3e 12-Hi cube at 1~TB/s takes 0.036 seconds, resulting in 27.78 TPS.
However, reading all data stored in a 576 GB HBF cube at 1~TB/s takes 0.576 seconds, resulting in 1.74 TPS.
From a service level objective (SLO) perspective, this level of TPS is unacceptable except for offline mode without SLO constraints.
Therefore, workloads suitable for HBF must exhibit sparse access, where only a portion of the stored large volume of data is accessed during the actual processing.

\begin{graybox}
\textit{\textbf{Motivation 3: All four conditions must be satisfied simultaneously to employ HBF effectively without performance or reliability degradation.}} 
\end{graybox}

Considering the four conditions above, \textit{large-scale MoE model inference with a low batch size} \textcolor{black}{can be one of the workloads suitable for HBF.}
First, as mentioned in Sec.~\ref{subsec:sota_llm_trends}, the latest frontier MoE models require hundreds of GB just for storing model weights and operate only in a read-only manner during inference.
Therefore, if HBF and other memory are configured heterogeneously and read-only model weights are stored in HBF (while other data is stored in memory outside HBF), the low write endurance problem can be avoided when processing large amounts of data.
Since the model weight access granularity used in large-scale MoE model inference is known to be larger than a 4 KB page unit~\cite{nam2026rome}, this also satisfies the above condition.
Finally, assuming low batch operation, only a subset of the total experts is activated for each input.

When large-scale MoE inference is deployed in data centers, a mismatch arises between HBF-suitable workload characteristics and conventional serving practices. 
Data center systems typically use large batch sizes to maximize resource utilization and reduce cost, which diminishes the effective sparsity of expert accesses. 
By contrast, PAICs are designed for single-user operation and typically run at low batch sizes, allowing MoE workloads to preserve sparse expert-access patterns while satisfying the other three HBF requirements. 
Therefore, integrating HBF into a PAIC not only addresses its memory-capacity and bandwidth limitations (Motivations 1 and 2) but also provides a deployment environment well suited to large-scale MoE inference (Motivation 3).
\textcolor{black}{This does not imply that HBF itself is unsuitable for data center systems. 
Rather, our argument is specific to low batch, large-scale MoE inference, where PAICs better preserve the sparse expert-access characteristics required to effectively exploit HBF.}

\section{HyPAIC: A Personal AI Computer with HBF}
\label{sec:paic_hbf}

\subsection{Hardware Architecture} 
\label{subsec:paic_hbf_hw_archi}

\begin{figure}[t]
    \centering
    \begin{subfigure}[t]{\linewidth}
        \centering
        \includegraphics[width=0.9\linewidth]{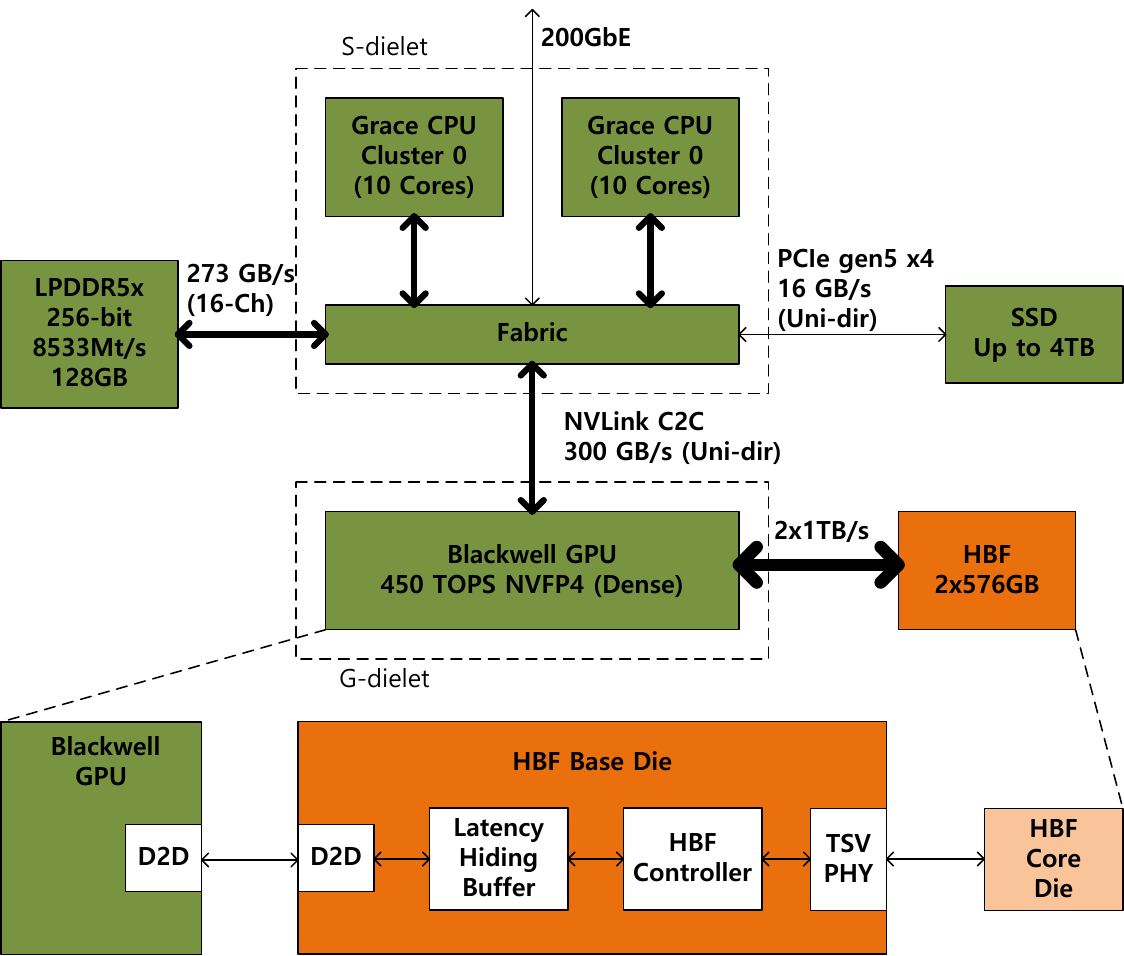}
        \caption{System architecture}
        \label{fig:paic_hbf_hw_archi:a}
        \vspace{1.0pc}
    \end{subfigure} \vfill
    \begin{subfigure}[t]{\linewidth}
        \centering
        \includegraphics[width=0.9\linewidth]{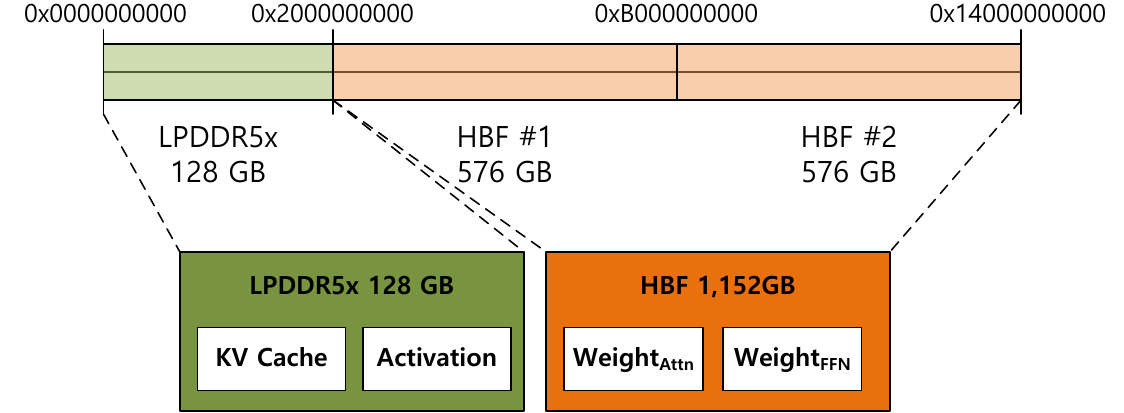}
        \caption{Unified address space and data placement}
        \label{fig:paic_hbf_hw_archi:b}
    \end{subfigure}      
   \begin{comment}
    \begin{subfigure}[t]{\linewidth}
        \centering
        \includegraphics[width=0.8\linewidth]{./figure/bw_saturation_v0.1.pdf}
        \caption{Minimum data size to saturate HBF bandwidth}
        \label{fig:paic_hbf_hw_archi:c}
    \end{subfigure}
    \end{comment}
    \caption{HyPAIC hardware architecture overview.}
    \vspace{-1pc}
    \label{fig:paic_hbf_hw_archi}
\end{figure}

\begin{figure}[t]
    \begin{subfigure}[t]{\linewidth}
        \centering
        \includegraphics[width=1.0\linewidth]{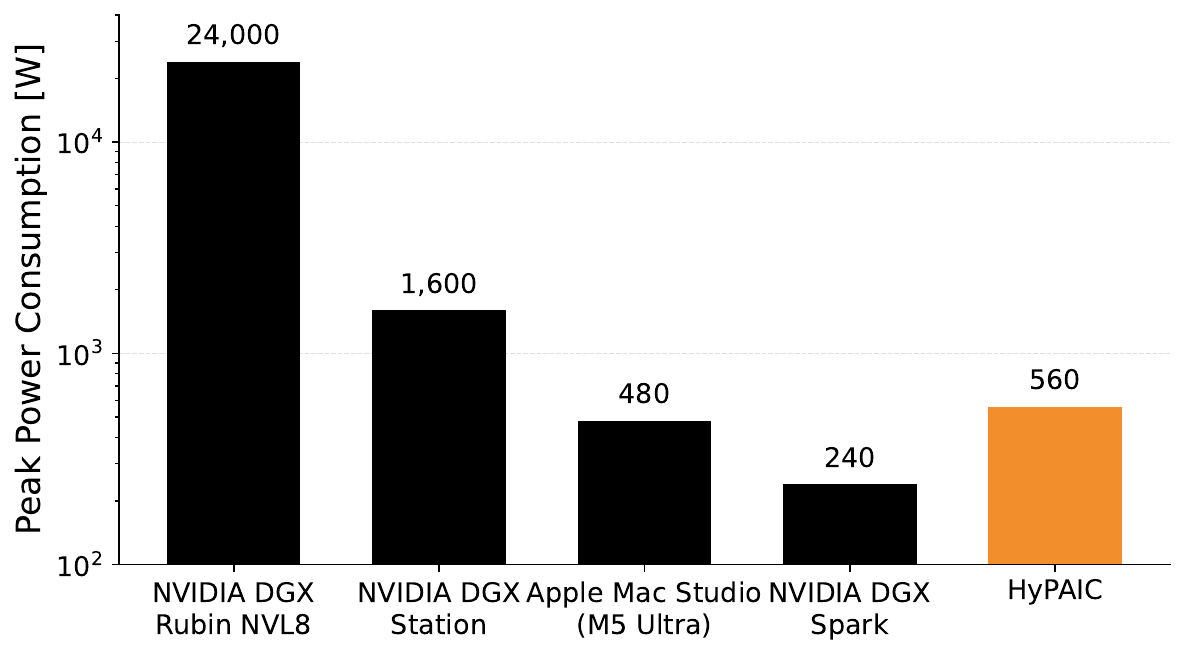}
    \end{subfigure}
    \caption{\textcolor{black}{Comparison of peak power between some AI PCs and AI workstations~\cite{mac-studio,nvl8,dgx-spark,dgx-station}.}}
    \label{fig:peak_power}
\end{figure}

% The proposed PAIC architecture incorporating HBF features a configuration where HBF is directly connected to the GPU, as shown in Fig.~\ref{fig:paic_hbf_hw_archi:a}.
% We propose a new PAIC architecture called \textbf{HyPAIC (Hyper-PAIC)}. 
% In this architecture, HBF is directly connected to the GPU die, as shown in Fig.~\ref{fig:paic_hbf_hw_archi:a}.
We propose a new PAIC architecture, \textbf{HyPAIC (Hyper-PAIC)}, wherein HBF is directly connected to the GPU die (Fig.~\ref{fig:paic_hbf_hw_archi:a}).
% Considering the GPU die size, it is assumed that two HBF cubes are installed.
% Since two HBF cubes are installed, the total capacity is 768 GB, and a bandwidth of 2 TB/s is supported (see Sec.~\ref{subsec:hbf}).
\textcolor{black}{GPU die and the HBF base die are connected via a die-to-die (D2D) interconnect such as UCIe, and 2.5D packaging technology, such as chip-on-wafer-on-substrate (CoWoS), is required to support high bandwidth~\cite{ocp-hbf-spec-v07}.}
Considering the GPU die size, we assume that two HBF cubes are integrated, which collectively provide a total capacity of 1,152 GB and support a bandwidth of 2 TB/s (see Sec.~\ref{subsec:hbf}).
In the detailed architecture of HyPAIC, an HBF controller is located within the base die, similar to custom HBM~\cite{kuemerle2025memory}, and a latency hiding buffer (LHB), proposed in~\cite{ha2026h}, is included to compensate for the high latency of NAND flash.
The capacity of the LHB is expected to be around 6~MB per HBF cube based on double buffering, assuming a read latency of 3~\si{\micro\second}, used in~\cite{kyung2026high,cheong2018flash}, and a 1~TB/s bandwidth.
We assume the TDP of each HBF cube to be 160 W, as in~\cite{ha2026h}.
This is 4$\times$ the TDP of HBM3e, which is the most conservative figure among existing studies~\cite{son2026exploring,wu2026memexplorer,ma2026challenges}.
\textcolor{black}{Reflecting the HBF's TDP, the maximum power consumption of the HyPAIC is 560 W, which is approximately 80 W higher than that of the latest high-performance PAIC, the Apple Mac Studio~\cite{mac-studio} (Fig.~\ref{fig:peak_power}).
However, this can be considered a reasonable trade-off for the more than twofold increase in memory capacity and bandwidth.
While the Mac Studio is designed with a height of 95 mm to facilitate heat dissipation, the existing DGX Spark (240 W) is designed with a height of 50.5 mm.
If sufficient space for heat dissipation—similar to that of the Mac Studio—is provided, the HyPAIC may provide sufficient thermal headroom, although detailed thermal and acoustic co-design remains necessary.}

\begin{figure}[t]
    \begin{subfigure}[t]{\linewidth}
        \centering
        \includegraphics[width=1.0\linewidth]{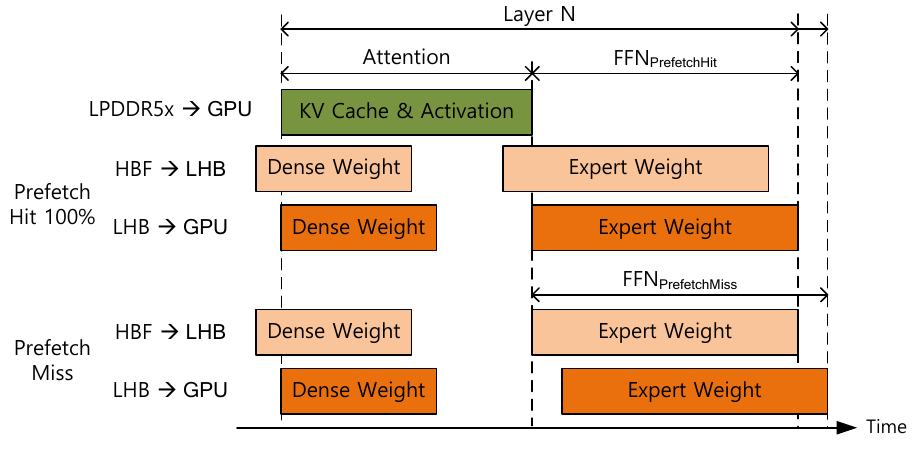}
    \end{subfigure}
    \caption{\textcolor{black}{Conceptual data-transfer flow in HyPAIC during MoE inference (Bar lengths are illustrative and not drawn to scale).}}
    \vspace{-1pc}
    \label{fig:time}
\end{figure}

In this architecture, both LPDDR5x and HBF operate as the GPU's main memory.
Therefore, the two memory types share a unified address space (Fig.~\ref{fig:paic_hbf_hw_archi:b}).
Data that is large in volume, read-only, coarse-grained, and sparsely accessed is stored in HBF, and mutable data is stored in LPDDR5x.
Since LPDDR5x and HBF are deployed on distinct physical interfaces and dielets, they can independently achieve their maximum bandwidth without direct memory-channel contention.
That is, the GPU can theoretically utilize the 273 GB/s bandwidth of LPDDR5x and the 2 TB/s bandwidth of HBF simultaneously.

There are also overheads in the HyPAIC architecture. 
Considering the area occupied by the HBF PHYs, the GPU computational performance is assumed to be 450 TOPS (NVFP4, dense), a 10\% decrease compared to that of the baseline, and power consumption increases due to the addition of HBF. 
Furthermore, the area overhead of a 6~MB LHB is 1.21\si{\milli\meter\squared} assuming 3\si{\nano\meter} SRAM~\cite{wu20223nm}, which is only about 1\% of the 121\si{\milli\meter\squared} base die area, as analyzed in~\cite{ha2026h}.
These overheads are acceptable if the capacity and performance benefits of HBF outweigh the associated area, compute, and power costs.

\subsection{Usage: Large-scale MoE Inference with low batch Size}
\label{subsec:paic_hbf_usage}

\begin{comment}
\begin{figure}[t]
  \centering
  \includegraphics[width=0.85\linewidth]{./figure/data_placement_v0.2.pdf} 
  \caption{HyPAIC data placement for large-scale MoE inference.}
  \label{fig:data_placement}
  \vspace{-1pc}
\end{figure}

\begin{figure}[t]
  \centering
  \includegraphics[width=1.0\linewidth]{./figure/decode_time_v0.2.pdf} 
  \caption{Comparison of decode time per token in Kimi~K2.5~\cite{team2026kimi} (ISL and OSL are both 32K, respectively).}
  \label{fig:decode_time}
  \vspace{-1pc} 
\end{figure}
\end{comment}

As discussed in Sec.~\ref{subsec:hbf_workload}, large-scale MoE inference at low batch sizes is well suited to HBF-equipped PAICs. 
HyPAIC stores KV caches and activations in the existing 128 GB of LPDDR5x while placing model weights in HBF (Fig.~\ref{fig:paic_hbf_hw_archi:b}). 
HBF is configured as a multi-channel system to maximize bandwidth utilization.
Assuming 16 channels per cube and 64 planes per channel, derived from the 16-plane/4-rank configuration in~\cite{maeda2026stco}, an aggregate access size of at least 4 MB is required to fully utilize the HBF bandwidth. % (Fig.~\ref{fig:paic_hbf_hw_archi:c}). 
Because the expert-weight tensors of recent MoE models are tens of megabytes in size~\cite{qwen3.5,team2026kimi}, they are sufficiently large to saturate the available HBF bandwidth.

The key advantage is that the GPU can utilize the bandwidth of both LPDDR5x and HBF to access the KV cache and model weights without contention \textcolor{black}{(Fig.~\ref{fig:time}).
During the attention process, both dense weights and the KV cache can be read simultaneously.
Since the dense weight values are fixed, they can be prefetched in advance.
As will be discussed later, it is also known that expert weights can be prefetched by predicting them with high accuracy~\cite{yu2026patterns,abts2022groq,sridharan2023chakra}.}
Another advantage is that storing all model weights in HBF frees up LPDDR5x capacity, allowing for the storage of more KV cache data. 
This enables the processing of larger batch sizes or longer contexts. 
While a larger batch size offers significant benefits to overall system throughput, it may not be favorable from the perspective of user experience, which we characterize using per-task decode throughput. 
In Sec.~\ref{subsec:ux}, we will present the results of an analysis on the scalability limits of the batch size.
% 추가: HBF 최소 데이터 접근 단위가 4MB 수준 (4KB * 1024) (IMW'26 논문 참고)

\begin{comment}
Fig.~\ref{fig:decode_time} shows a comparison of the decode time per token of Kimi~K2.5~\cite{team2026kimi} depending on whether HBF is adopted.
To isolate the performance benefits of HBF's superior bandwidth from its capacity advantages, we temporarily evaluate an idealized, capacity-unconstrained LPDDR5x baseline that can fully accommodate the model weights.
Simulation-based experimental results show that the decode time per token is reduced by 9.24 $\times$ due to the advantage that HBF itself supports high bandwidth, enabling the HyPAIC to utilize HBF and LPDDR5x simultaneously.
Thus, even assuming no SSD access, the decode processing speed can be faster due to the application of HBF.
Since decoding is completely memory-bound, sacrificing 10\% of GPU computational performance has almost no impact on overall performance; rather, the bandwidth benefits outweigh the compute-performance penalty.
\end{comment}

\subsection{Software Support}
\label{subsec:paic_hbf_sw_support}

\begin{figure}[t]
    \begin{subfigure}[t]{\linewidth}
        \centering
        \includegraphics[width=0.9\linewidth]{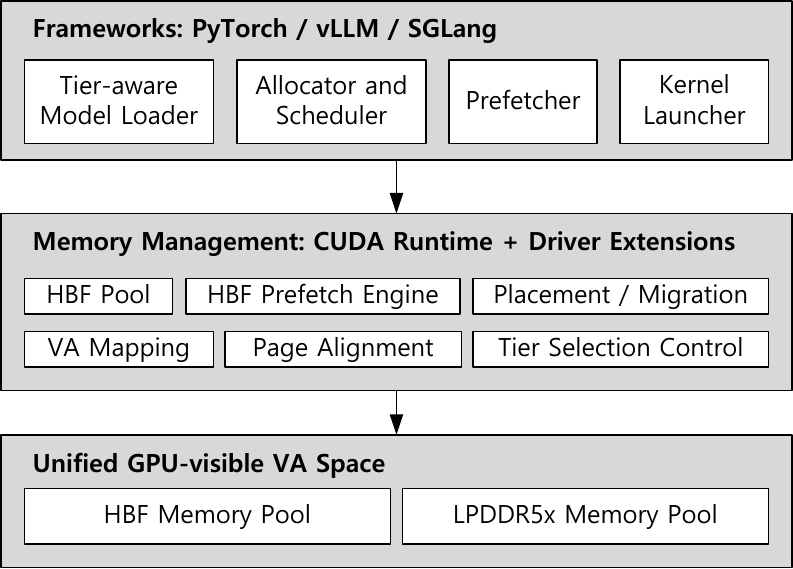}
    \end{subfigure}
    \caption{Software support for HyPAIC.}
    \vspace{-1pc}
    \label{fig:sw_support}
\end{figure}

To use HyPAIC effectively, software support at multiple layers is essential (Fig.~\ref{fig:sw_support}).

\noindent \textbf{Memory Management: }
HyPAIC exposes LPDDR5x and HBF through a unified GPU-visible virtual address space but requires tier-aware memory placement to exploit their distinct characteristics. 
Existing CUDA systems provide several useful mechanisms~\cite{cuda}, although none can be used for HBF without extension. 
Device and host allocation APIs, such as  \texttt{cudaMalloc} and \texttt{cudaMallocHost}, allocate device-resident and page-locked host-resident memory, respectively, but do not natively distinguish multiple memory tiers attached to the same GPU. 
For managed or supported system-allocated memory, \texttt{cudaMemAdvise} can provide preferred-location advice, while \texttt{cudaMemPrefetchAsync} can explicitly request asynchronous migration to a specified location, but neither guarantees permanent residency at that location.

Coherent systems such as GH200~\cite{gh200} and GB200~\cite{gb200} provide a relevant precedent. 
Their CPU LPDDR and GPU HBM are accessible through a unified address space. 
Under operating modes in which GPU memory is exposed as a NUMA memory node, applications can influence the placement of OS-managed pages using \texttt{numactl}, \texttt{libnuma}, or \texttt{mbind}. 
This NUMA-based mechanism is implementation-dependent because some driver modes manage GPU memory internally without exposing it as an OS NUMA node.

HyPAIC can build on these CUDA unified memory, virtual-memory, and NUMA mechanisms but requires additional runtime and driver support. 
In particular, HBF must be exposed as a distinct allocatable memory location or memory pool so that read-only model weights can be explicitly mapped to HBF while mutable KV caches, activations, and temporary buffers remain in LPDDR5x. 
The runtime must also support HBF-page-aligned allocation and size rounding, strict tier-placement or migration control, and HBF-specific prefetch hints. 
Thus, HyPAIC does not require an entirely new memory-management abstraction but does require tier-aware extensions to existing CUDA and operating-system mechanisms.

\begin{comment}
In the HyPAIC architecture, LPDDR5x and HBF serve as unified main memory, as discussed in Sec.~\ref{subsec:paic_hbf_hw_archi}.
Therefore, memory management must be supported to efficiently utilize heterogeneous memory.
Fortunately, NVIDIA CUDA supports this heterogeneous memory management~\cite{cuda}.
This is already being used in architectures such as the GH200~\cite{gh200} and GB200~\cite{gb200}, which are equipped with LPDDR and HBM.
The first method is to use CUDA APIs.
Developers can specify the physical location of the data through explicit API calls (e.g., \texttt{cudaMalloc}/\texttt{cudaMallocHost}), or they can specify a preferred location (e.g., \texttt{cudaMemAdvise}).
The second method is to use NUMA APIs.
Architectures like the GH200 or GB200 recognize the CPU and LPDDR as one NUMA node, and the GPU and HBM as another NUMA node.
Therefore, developers can utilize NUMA APIs to ensure data is stored in a specific NUMA node (e.g., \texttt{numactl}/\texttt{mbind}).
HyPAIC cannot use these mechanisms unchanged.
Modifications will be necessary to accommodate HBF, but the point is that the extent of such modifications will not be significant.
\end{comment}

\begin{figure}[t]
    \begin{subfigure}[t]{\linewidth}
        \centering
        \includegraphics[width=0.9\linewidth]{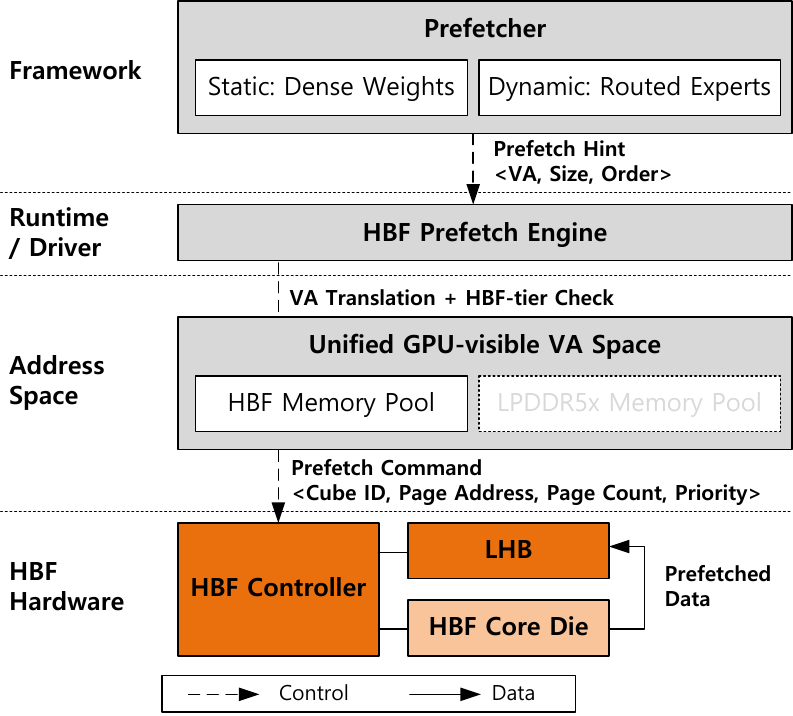}
    \end{subfigure}
    \caption{Prefetch flow for HyPAIC.}
    \vspace{-1pc}
    \label{fig:prefetch}
\end{figure}

\noindent \textbf{Deep Learning Frameworks: }
Once the CUDA runtime exposes HBF as a distinct memory pool, LLM inference frameworks such as PyTorch~\cite{paszke2019pytorch}, vLLM~\cite{kwon2025vllm}, and SGLang~\cite{zheng2024sglang} require tier-aware allocation and scheduling support. 
A tier-aware model loader can allocate immutable model parameters—including attention, embedding, and expert weights—in HBF while placing mutable runtime state, such as KV caches, activations, routing buffers, and temporary workspaces, in LPDDR5x. 
Existing allocator and virtual-memory-pool mechanisms provide relevant framework-level extension points, although they require underlying driver support for selecting HBF as a physical memory tier.

The framework must also coordinate HBF prefetching with model execution (Fig.~\ref{fig:prefetch}). 
Once the model has been loaded, the addresses and relative access order of dense model parameters remain invariant across inference iterations and can therefore be converted into a static, operation-ordered prefetch schedule~\cite{abts2022groq,sridharan2023chakra}. 
In contrast, routed-expert accesses depend on runtime router decisions and require dynamic prefetch hints.
Prior studies have shown that \textcolor{black}{1) there are expert correlations between prefill and decode phases and between layers~\cite{yu2026patterns} and 2)} future expert activations can be predicted with useful accuracy~\cite{zhu2025pre,hwang2024pre,fang2025accurate}; HyPAIC can use these predictors to issue speculative prefetch requests. 
The framework associates each request with the relevant execution stream and provides the expert-weight address range, size, and expected access order. 
The runtime and driver translate these hints into HBF prefetch commands, while the HBF controller stages the requested pages into the LHB and manages their replacement. 
Mispredictions may expose HBF latency or waste HBF bandwidth and LHB capacity.

\section{Evaluation}
\label{sec:eval}

\subsection{Experimental Setup}
\label{subsect:exp_setup}

\noindent \textbf{Methods: }
We evaluate the throughput, TDP-normalized throughput, and task concurrency of baseline DGX Spark and HyPAIC using \textcolor{black}{an architecture-level simulator.
For our architecture-level simulation, we have extended NVIDIA AIConfigurator~\cite{xu2026aiconfigurator}.
We add DGX Spark and HyPAIC system models to AIConfigurator v0.11.0 and implement a heterogeneous-memory operation mechanism where the KV cache is stored in LPDDR5x and weights are stored in HBF.
All other hardware operation models are inherited unchanged from AIConfigurator.}
We assume that HBF provides HBM3e-level bandwidth and that LHB prefetching operates seamlessly.
However, for an in-depth analysis, we also examine performance variations as a function of HBF bandwidth (Sec.~\ref{subsec:bw_sensitivity}) and LHB prefetching hit rate (Sec.~\ref{subsec:prefetch_miss}).

\noindent \textbf{Workloads: }
We evaluate the latest open-source MoE models (Qwen3.5~\cite{qwen3.5} and Kimi~K2.5~\cite{team2026kimi}).
For Qwen3.5, we select the 35B-A3B model, which can be loaded into main memory of a single PAIC (128 GB LPDDR5x), and the 397B-A17B model, which cannot be loaded onto a single PAIC's main memory.
We also select the Kimi~K2.5 model, which can represent a trillion-parameter MoE model.

\noindent \textbf{System Configurations:}
We configure the baseline system as the latest DGX Spark~\cite{dgx-spark} (Fig.~\ref{fig:paic_baseline} and Table~\ref{tab:paics}).
In configurations with multiple DGX Sparks, intermediate results are transmitted via 200 GbE.
For multiple DGX Sparks, we select the parallelism combination that shows the best performance.
For Qwen3.5-397B-A17B and Kimi~K2.5, 2 and \textcolor{black}{8} DGX Sparks are required, respectively.
We do not evaluate 2 DGX Sparks for Qwen3.5-35B-A3B, and 4 DGX Sparks for Qwen3.5-397B-A17B.

The proposed HyPAIC has two HBF cubes connected to the GPU die, with each HBF cube supporting 576 GB and 1 TB/s.
It is assumed that the GPU's computational performance is reduced by 10\% due to the inclusion of HBF PHYs (Fig.~\ref{fig:paic_hbf_hw_archi}).

\begin{figure*}[!t]  
    \centering
    
    \begin{subfigure}[!t]{0.32\textwidth}
        \centering
        \includegraphics[width=1.0\linewidth]{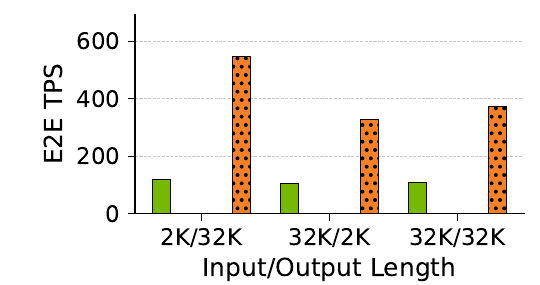}
        \caption{Qwen3.5-35B-A3B}
        \label{fig:qwen35-1col}
    \end{subfigure}
    \hfill    
    \begin{subfigure}[!t]{0.32\textwidth}
        \centering
        \includegraphics[width=1.0\linewidth]{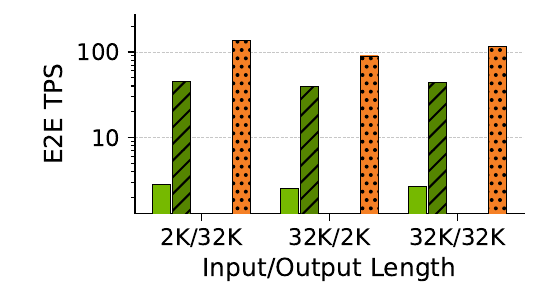}
        \caption{Qwen3.5-397B-A17B}
        \label{fig:qwen397-1col}
    \end{subfigure}
    \hfill
    \begin{subfigure}[!t]{0.32\textwidth}
        \centering
        \includegraphics[width=1.0\linewidth]{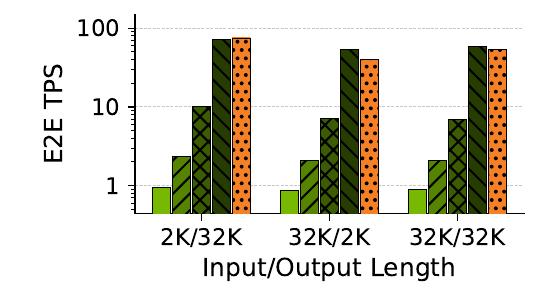}
        \caption{Kimi~K2.5}
        \label{fig:kimi-1col}
    \end{subfigure}
    
    \vspace{-0.5pc}   
    
    \begin{center}
        \includegraphics[width=0.6\linewidth]{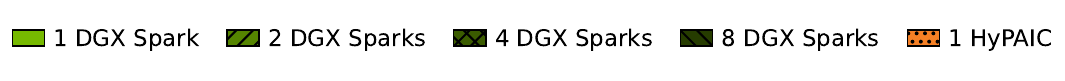}
    \end{center}
    
    \vspace{-1pc}   
    
    \caption{\textcolor{black}{E2E system throughput (Higher is better).}}

    \label{fig:system_throughput}
\end{figure*}

\subsection{System Throughput}
\label{subsec:system_throughput}

We evaluate E2E system throughput for various ISL and OSL combinations at a batch size of one (Fig.~\ref{fig:system_throughput}).

\noindent \textbf{Qwen3.5-35B-A3B: }
The baseline always achieves over 100 E2E TPS.
However, if HBF is applied, a maximum of \textcolor{black}{546.47} E2E TPS can be achieved.
This is because HBF itself supports higher bandwidth compared to LPDDR5x, and HBF can be used simultaneously with LPDDR5x.

\noindent \textbf{Qwen3.5-397B-A17B: }
Because the size of this model exceeds the LPDDR5x capacity of a single DGX Spark, SSD access occurs, and as a result, 10 E2E TPS cannot be achieved for 1 DGX Spark.
However, by utilizing 2 DGX Sparks, it is possible to achieve \textcolor{black}{tens of E2E TPS, but fewer than 50 E2E TPS} in all three sequence length combinations.
This means that doubling the capacity and bandwidth of LPDDR5x has a greater impact than using 200 GbE to transmit activations.
With HyPAIC, it is possible to achieve \textcolor{black}{almost} 100 E2E TPS in all three sequence length combinations, even though only one PAIC is used.

\noindent \textbf{Kimi~K2.5: }
In the case of this model, it fails to achieve 10 E2E TPS in all three sequence length combinations, even with 2 DGX Sparks.
% In the case of 4 DGX Sparks, approximately 50 E2E TPS can be achieved in all three sequence length combinations.
\textcolor{black}{Even when using 4 DGX Sparks, the system exceeds 10 E2E TPS only in the 2K/32K case, while showing lower performance in the other scenarios.
8 DGX Sparks must be used to achieve performance exceeding 50 E2E TPS in all scenarios.}
However, in the case of HyPAIC, a single HyPAIC approximately matches or exceeds the throughput of \textcolor{black}{8} DGX Spark systems. % (up to 1.87$\times$ improvement at 2K/32K).

Because HyPAIC supplements LPDDR5x with higher-bandwidth HBF, its performance advantage grows as the decode phase accounts for a larger fraction of execution time. 
\textcolor{black}{In other words, the benefits of HBF may be reduced when the input sequence length, which has a greater impact on the prefill phase, is long.
Combined with HyPAIC’s assumed 10\% reduction in compute performance and increased number of baseline DGX Sparks, these cause HyPAIC’s system throughput to be 25.25\% and 8.81\% lower than that of the 8 DGX Sparks baseline for 32K/2K and 32K/32K, respectively.
Notably, however, HyPAIC can deliver sufficient performance for use with just a single device.}

\subsection{TDP-normalized Throughput}

\begin{figure*}[!t]  
    \centering
    
    \begin{subfigure}[t]{0.32\textwidth}
        \centering
        \includegraphics[width=1.0\linewidth]{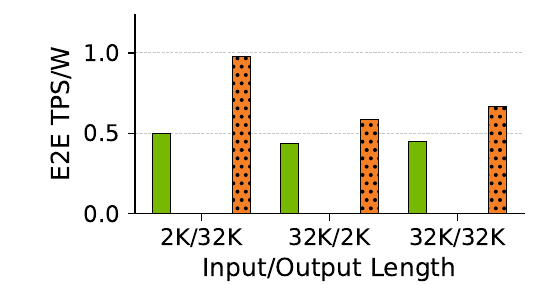}
        \caption{Qwen3.5-35B-A3B}
        \label{fig:power_efficiency_a}
    \end{subfigure}
    \hfill   
    \begin{subfigure}[t]{0.32\textwidth}
        \centering
        \includegraphics[width=1.0\linewidth]{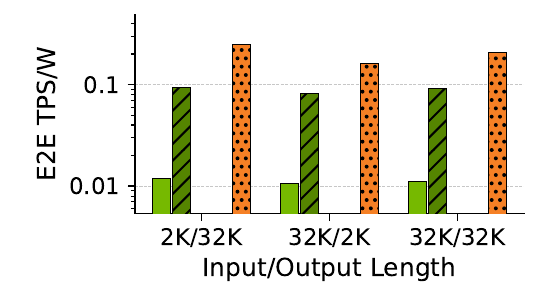}
        \caption{Qwen3.5-397B-A17B}
        \label{fig:power_efficiency_b}
    \end{subfigure}
    \hfill
    \begin{subfigure}[t]{0.32\textwidth}
        \centering
        \includegraphics[width=1.0\linewidth]{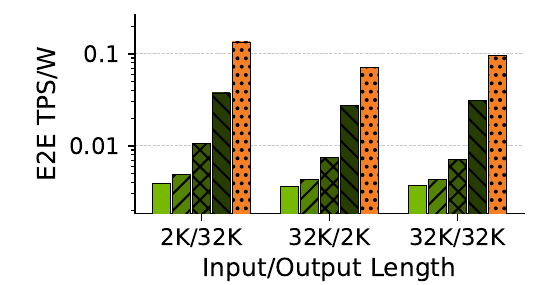}
        \caption{Kimi~K2.5}
        \label{fig:power_efficiency_c}
    \end{subfigure}
    
    \vspace{-0.5pc}   
    
    \begin{center}
        \includegraphics[width=0.6\linewidth]{./figure/unified_legend_all_metrics_v1.1.pdf}
    \end{center}
    
    \vspace{-1pc}   
    
    \caption{\textcolor{black}{E2E TDP-normalized throughput (Higher is better).}}
    \vspace{-1pc}
    \label{fig:power_efficiency}
\end{figure*}

Since HyPAIC is the architecture that adds HBF to the existing PAIC, it naturally shows better performance \textcolor{black}{in many cases}.
For a fairer comparison, we also carry out a TDP-normalized throughput comparison by dividing E2E TPS by the total TDP (Fig.~\ref{fig:power_efficiency}).
Since we assumed the two HBF cubes consume 320~W in total, HyPAIC’s TDP-normalized throughput advantage is smaller than its throughput advantage when it is compared with configurations containing at most 2 DGX Sparks.
However, compared to the Kimi~K2.5 baseline which requires \textcolor{black}{8} DGX Sparks, TDP-normalized throughput is improved by up to \textcolor{black}{3.59}$\times$.
Even at 32K/2K, where there is no significant performance \textcolor{black}{improvement} in E2E TPS, E2E TDP-normalized throughput is improved by \textcolor{black}{2.56}$\times$.

\begin{table}[!t]
\caption{\textcolor{black}{Maximum task concurrency maintaining 50 decode TPS per task}}
\vspace{-0.5pc}
\begin{center} 
\begin{tabularx}{\columnwidth}{
|c|
c|
>{\centering\arraybackslash}X|
>{\centering\arraybackslash}X|
>{\centering\arraybackslash}X|
>{\centering\arraybackslash}X|
c|
}
\hline
\multirow{2}{*}{\textbf{Models}} &
\multirow{2}{*}{\textbf{ISL/OSL}} &
\multicolumn{4}{c|}{\textbf{Number of DGX Sparks}} &
\multirow{2}{*}{\textbf{1 HyPAIC}} \\
\cline{3-6}
& & \textbf{1} & \textbf{2} & \textbf{4} & \textbf{8} & \\
\hline

% ============================================
% Qwen3.5-35B-A3B
% ============================================
\multirow{3}{*}{
\textbf{
\begin{tabular}{@{}c@{}}
Qwen3.5\\
-35B-A3B\\
\cite{qwen3.5}
\end{tabular}
}}
& 2K/32K
& 4
& \multirow{3}{*}{-}
& \multirow{3}{*}{-}
& \multirow{3}{*}{-}
& 16 \\
\cline{2-3}\cline{7-7}

& 32K/2K
& 3
&
&
&
& 12 \\
\cline{2-3}\cline{7-7}

& 32K/32K
& 3
&
&
&
& 9 \\
\hline

% ============================================
% Qwen3.5-397B-A17B
% ============================================
\multirow{3}{*}{
\textbf{
\begin{tabular}{@{}c@{}}
Qwen3.5\\
-397B-A17B\\
\cite{qwen3.5}
\end{tabular}
}}
& 2K/32K
& \multirow{3}{*}{-}
& \multirow{3}{*}{0}
& \multirow{3}{*}{-}
& \multirow{3}{*}{-}
& \multirow{3}{*}{3} \\
\cline{2-2}

& 32K/2K
&
&
&
&
& \\
\cline{2-2}

& 32K/32K
&
&
&
&
& \\
\hline

% ============================================
% Kimi K2.5
% ============================================
\multirow{3}{*}{
\textbf{
\begin{tabular}{@{}c@{}}
Kimi~K2.5\\
\cite{team2026kimi}
\end{tabular}
}}
& 2K/32K
& \multirow{3}{*}{-}
& \multirow{3}{*}{-}
& \multirow{3}{*}{0}
& \multirow{3}{*}{1}
& \multirow{3}{*}{1} \\
\cline{2-2}

& 32K/2K
&
&
&
&
& \\
\cline{2-2}

& 32K/32K
&
&
&
&
& \\
\hline

\end{tabularx}
\end{center}
\label{tab:task_concurrency}
\vspace{-1pc}
\end{table}

\subsection{User Experience}
\label{subsec:ux}

While E2E system throughput and TDP-normalized throughput are related to the operating costs of the LLM serving system, user experience is related to user satisfaction through the task processing speed.
Even if a model is handling difficult tasks, user satisfaction is bound to decrease if it processes them at a very slow speed (e.g., 1 decode TPS per task).
Although the target processing speed may vary depending on the type of workload, recent studies suggest that at least 50 decode TPS per task is desirable~\cite{50tps,50tps2}.

We examine how many tasks can be run simultaneously for three models while maintaining 50 decode TPS per task, which is directly related to user productivity (Table~\ref{tab:task_concurrency}).

\noindent \textbf{Qwen3.5-35B-A3B: }
When running Qwen3.5-35B-A3B on 1 DGX Spark, it is possible to run up to \textcolor{black}{4} tasks simultaneously while maintaining 50 decode TPS per task.
However, with HyPAIC, it is possible to maintain 50 decode TPS per task while running up to \textcolor{black}{16} tasks simultaneously.
When \textcolor{black}{16} tasks are run simultaneously, \textcolor{black}{selecting multiple experts can dilute the sparse access characteristic of MoE.
However, due to the very small model size of Qwen3.5-35B-A3B, it is possible to achieve 50 decode TPS per task even when loading it into HBF and reading a large portion of the model.}
Namely, for Qwen3.5-35B-A3B, HBF capacity utilization is low, so reading the entire model is still equivalent to sparse access.

\noindent \textbf{Qwen3.5-397B-A17B \& Kimi~K2.5: }
For the larger models, \textcolor{black}{to maintain 50 decode TPS per task, there are cases where only a single task can be run, or where it is impossible to run even one task.
However, using HyPAIC enables the simultaneous processing of three tasks at over 50 decode TPS per task when running Qwen3.5-397B-A17B.
In addition, Kimi K2.5 is still limited to processing only one task at over 50 decode TPS per task.}
This is related to the sparse access discussed in Sec.~\ref{subsec:hbf_workload}.
Kimi~K2.5 is a 1T-parameter model and occupies a large portion of HyPAIC's HBF capacity.
When multiple experts are selected to process a large number of tasks simultaneously, the ratio of bandwidth to active capacity decreases, making it hard to maintain 50 decode TPS per task.

\begin{figure*}[t]  % [t] = 페이지 상단, [b] = 하단, [p] = 단독 페이지
    \centering
    
    % 첫 번째 행
    \begin{subfigure}[b]{0.32\textwidth}
        \centering
        \includegraphics[width=\linewidth]{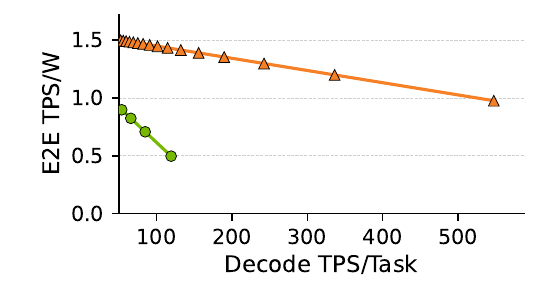}
        \caption{Qwen3.5-35B-A3B 2K/32K}
        \label{fig:sub1}
    \end{subfigure}
    \hfill  % 열 간격 균등 분배
    \begin{subfigure}[b]{0.32\textwidth}
        \centering
        \includegraphics[width=\linewidth]{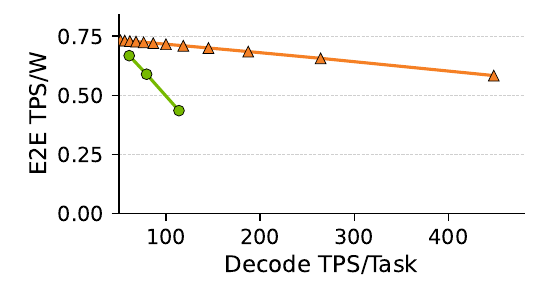}
        \caption{Qwen3.5-35B-A3B 32K/2K}
        \label{fig:sub2}
    \end{subfigure}
    \hfill
    \begin{subfigure}[b]{0.32\textwidth}
        \centering
        \includegraphics[width=\linewidth]{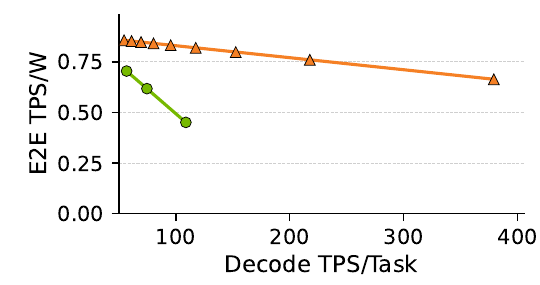}
        \caption{Qwen3.5-35B-A3B 32K/32K}
        \label{fig:sub3}
    \end{subfigure}

    \vspace{0.5pc}
    
    % 두 번째 행
    \begin{subfigure}[b]{0.32\textwidth}
        \centering
        \includegraphics[width=\linewidth]{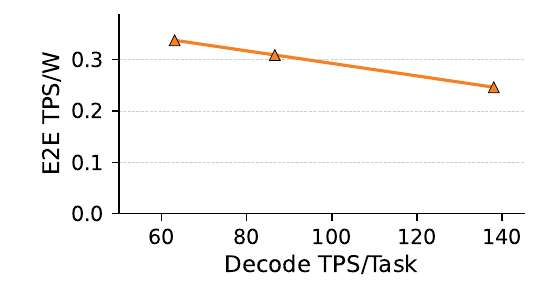}
        \caption{Qwen3.5-397B-A17B 2K/32K}
        \label{fig:sub4}
    \end{subfigure}
    \hfill
    \begin{subfigure}[b]{0.32\textwidth}
        \centering
        \includegraphics[width=\linewidth]{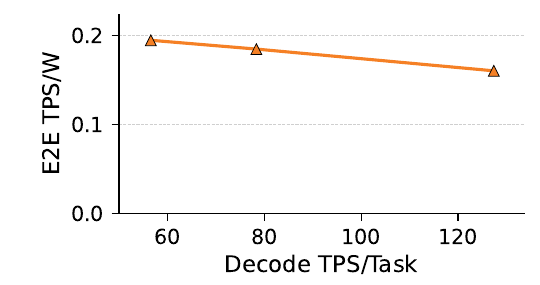}
        \caption{Qwen3.5-397B-A17B 32K/2K}
        \label{fig:sub5}
    \end{subfigure}
    \hfill
    \begin{subfigure}[b]{0.32\textwidth}
        \centering
        \includegraphics[width=\linewidth]{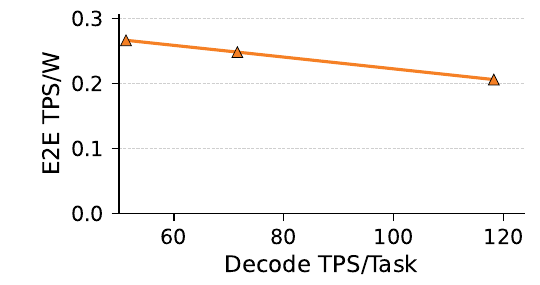}
        \caption{Qwen3.5-397B-A17B 32K/32K}
        \label{fig:sub6}
    \end{subfigure}

    \vspace{0.5pc}

\begin{comment}
    % 세 번째 행
    \begin{subfigure}[b]{0.32\textwidth}
        \centering
        \includegraphics[width=\linewidth]{./figure/kimi-K2.5_1t_a32b_2k32k_v0.3.pdf}
        \caption{Kimi~K2.5 2K/32K}
        \label{fig:sub4}
    \end{subfigure}
    \hfill
    \begin{subfigure}[b]{0.32\textwidth}
        \centering
        \includegraphics[width=\linewidth]{./figure/kimi-K2.5_1t_a32b_32k2k_v0.3.pdf}
        \caption{Kimi~K2.5 32K/2K}
        \label{fig:sub5}
    \end{subfigure}
    \hfill
    \begin{subfigure}[b]{0.32\textwidth}
        \centering
        \includegraphics[width=\linewidth]{./figure/kimi-K2.5_1t_a32b_32k32k_v0.3.pdf}
        \caption{Kimi~K2.5 32K/32K}
        \label{fig:sub6}
    \end{subfigure}
\end{comment}

    \begin{center}
        \includegraphics[width=0.4\linewidth]{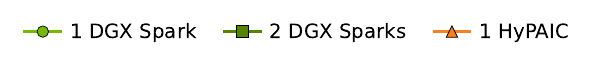}
    \end{center}

    \vspace{-1.0pc}

    \caption{\textcolor{black}{Relationship between decode throughput per task and E2E TDP-normalized throughput.}}
    \label{fig:pareto}
\end{figure*}

\subsection{TDP-normalized Throughput vs. Decode TPS per Task}
\label{subsec:pareto}

The results from~\ref{subsec:system_throughput} to~\ref{subsec:ux} demonstrate that HyPAIC can achieve better E2E TDP-normalized throughput compared to the baseline at the same decode TPS per task.
As shown in Fig.~\ref{fig:pareto}, HyPAIC's Pareto frontier is superior to the baseline in all cases.
\textcolor{black}{Qwen3.5-397B-A17B running on 2 DGX Sparks is omitted from Fig.~\ref{fig:pareto} because it cannot achieve 50 decode TPS per task.
In the case of Kimi K2.5, both the 8 DGX Sparks and 1 HyPAIC configurations can run only a single task to maintain a performance of at least 50 decode TPS per task; consequently, no Pareto frontier exists, and they are excluded from Fig.~\ref{fig:pareto}.
In the case of Qwen3.5-35B-A3B, using HyPAIC delivers up to 2.88$\times$ higher E2E TPS/W while supporting comparable decode TPS.}

\begin{comment}
For Kimi~K2.5, E2E TDP-normalized throughput remains nearly constant across the task concurrency because bandwidth for experts is already saturated at low concurrency. 
Increasing task concurrency therefore primarily redistributes the nearly fixed aggregate throughput among concurrent tasks, reducing decode TPS per task without substantially increasing total throughput. 
Since the system TDP is fixed for each hardware configuration, its E2E TPS/W also remains nearly constant.
\end{comment}

\subsection{Bandwidth Sensitivity}
\label{subsec:bw_sensitivity}

\begin{figure}[!t]  
    \centering
    
    \begin{subfigure}[t]{\linewidth}
        \centering
        \includegraphics[width=0.8\linewidth]{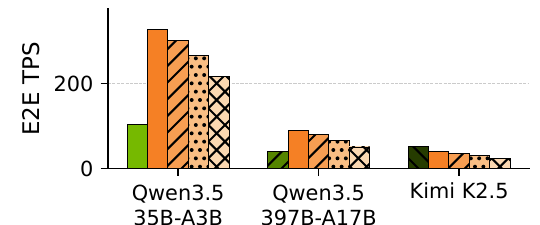}
        \caption{System throughput}
        \label{fig:bw_sensitivity_tps_tpsw_a}
    \end{subfigure} 
    
    \begin{subfigure}[t]{\linewidth}
        \centering
        \includegraphics[width=0.8\linewidth]{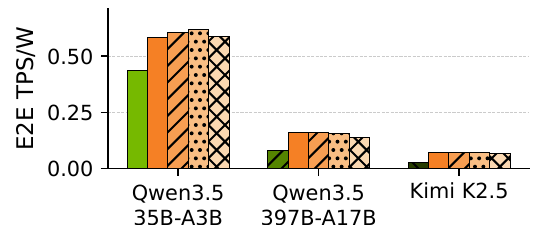}
        \caption{TDP-normalized throughput}
        \label{fig:bw_sensitivity_tps_tpsw_b}
    \end{subfigure}

    \begin{center}
        \includegraphics[width=1.0\linewidth]{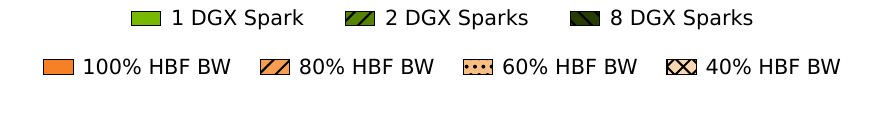}
    \end{center}    
    \vspace{-1pc}
    \caption{\textcolor{black}{System throughput and TDP-normalized throughput of 32K/2K case inference according to HBF bandwidth change.}}
    \vspace{-1pc}
    \label{fig:bw_sensitivity_tps_tpsw}
\end{figure}

HBF aims to support bandwidth at the level of HBM, but during the implementation process, it may end up supporting bandwidth that falls short of that~\cite{hbf2025}.
Therefore, we conduct an additional analysis on how performance is affected when the bandwidth of HBF is reduced.
We explore system throughput and TDP-normalized throughput while gradually reducing the HBF bandwidth.
% With the exception of the 32K/2K system throughput result, which initially shows lower performance even with 100\% HBF bandwidth supported, most results show a performance advantage over the baseline even when HBF bandwidth is reduced to a certain level.
% However, experimental results also indicate that if bandwidth below a certain level is supported (e.g., 40\% HBF bandwidth), the performance advantage over the baseline disappears.
\textcolor{black}{We carry out TPS and TPS/W experiments across three models using the 32K/2K configuration—the scenario where the benefits of HyPAIC are least pronounced—while progressively reducing the HBF bandwidth.
Note that HBF's TDP is reduced at the same rate as its bandwidth.
The results show that, with the exception of the Kimi K2.5 TPS, which exhibited inferior performance from the outset, all other cases demonstrated improved performance even when the HBF bandwidth was reduced to 40\% (Fig.~\ref{fig:bw_sensitivity_tps_tpsw}).
An interesting observation is that reducing the bandwidth to 80\% actually resulted in an increase in TPS/W across all three cases.
This suggests that an intermediate HBF bandwidth may offer a better power-efficiency trade-off.}

\subsection{Effect of LHB Prefetching Hit Rate}
\label{subsec:prefetch_miss}

\begin{figure}[t]
    \begin{subfigure}[t]{\linewidth}
        \centering
        \includegraphics[width=0.9\linewidth]{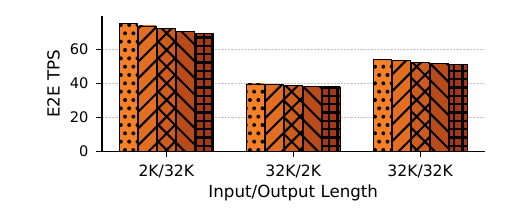}
    \end{subfigure}
    
    \vspace{-0.5pc}   
    
    \begin{center}
        \includegraphics[width=0.70\linewidth]{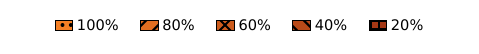}
    \end{center}

    \vspace{-1pc}

    \caption{\textcolor{black}{E2E system throughput of Kimi~K2.5 inference according to the LHB prefetch hit rate change.}}
    \label{fig:prefetch_miss}
\end{figure}

Although previous studies indicate that prefetching to LHB is possible because the next activated experts can be predicted with high accuracy~\cite{zhu2025pre,hwang2024pre,fang2025accurate}, we additionally analyze performance changes based on the prefetch hit rate.
We assume that the HBF read latency is 3~\si{\micro\second}, as mentioned in Sec.~\ref{subsec:paic_hbf_hw_archi}.
\textcolor{black}{In the case of a prefetch miss, it is assumed that a penalty occurs at the expert weight tensor level.
In other words, we assume that even if an unnecessary expert is prefetched due to an incorrect prediction—requiring the originally needed weight to be selected subsequently—the read latency penalty is incurred once for that expert; this is because, once an expert is selected, the subsequent data is read in a burst.}
Even though the expert prefetch hit rate is only 20\%, \textcolor{black}{performance degrades by at most 8.12\% (2K/32K case, Fig.~\ref{fig:prefetch_miss}).}
% This result shows that, except for 32K/2K where performance is lower than the baseline even at 100\% prefetch hit rate, superior performance to the baseline can be achieved in the remaining cases.
\textcolor{black}{These results indicate that, under our latency model, prefetch misses have only a limited impact on overall performance.}

\subsection{\textcolor{black}{Comparison with HBM-HBF Hybrid Architecture}}
\label{subsec:comparison_h3}

\begin{table}[t]
\caption{\textcolor{black}{Comparison of E2E TPS and E2E TPS/W for HyPAIC and PAIC with H\textsuperscript{3}}}
\vspace{-0.5pc}
\begin{center}
\begin{tabular}{|c|c|c|c|}
\hline
System                        & ISL/OSL & E2E TPS & E2E TPS/W \\ \hline
\multirow{3}{*}{HyPAIC}       & 2K/32K  & 75.31   & 0.134     \\ \cline{2-4} 
                              & 32K/2K  & 39.64   & 0.071     \\ \cline{2-4} 
                              & 32K/32K & 54.06   & 0.097     \\ \hline
\multirow{3}{*}{PAIC with H\textsuperscript{3}} & 2K/32K  & 77.07   & 0.120     \\ \cline{2-4} 
                              & 32K/2K  & 42.92   & 0.067     \\ \cline{2-4} 
                              & 32K/32K & 56.66   & 0.089     \\ \hline
\end{tabular}
\end{center}
\label{tab:h3}
\vspace{-1pc}
\end{table}

\textcolor{black}{This paper considers adding only HBF to the GPU die, taking practical implementation into account. 
However, it is also possible to adopt an architecture such as H\textsuperscript{3}~\cite{ha2026h}, which requires more advanced packaging technologies than HyPAIC.
In this case, the KV cache is stored in HBM, while the weights are stored in HBF. 
Assuming single-batch operation, the HBM capacity is sufficient.
As shown in Table~\ref{tab:h3}, while PAIC with H\textsuperscript{3} achieves up to 1.08 $\times$ higher E2E TPS, its E2E TPS/W is up to 10.46\% lower. 
Since each approach has its own advantages and disadvantages, these results highlight a trade-off: H\textsuperscript{3} provides slightly higher throughput, whereas HyPAIC offers higher power efficiency and lower implementation complexity.}

\subsection{\textcolor{black}{Comparison with Multi-PAIC supporting High Bandwidth Scale-out Interconnects}}
\label{subsec:scale-out_comparison}

\begin{figure}[!t]  
    \centering
    
    \begin{subfigure}[t]{\linewidth}
        \centering
        \includegraphics[width=0.9\linewidth]{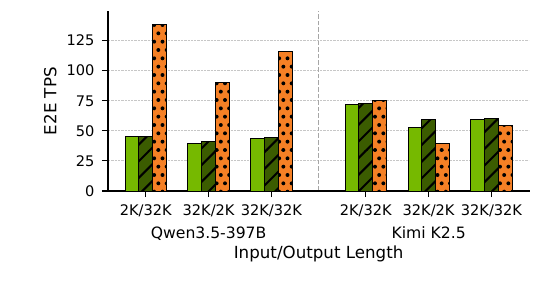}
        \vspace{-1pc}
        \caption{System throughput}
        \label{fig:scaleout_bw_e2e_tps_per_w_a}
    \end{subfigure} 
    
    \begin{subfigure}[t]{\linewidth}
        \centering
        \includegraphics[width=0.9\linewidth]{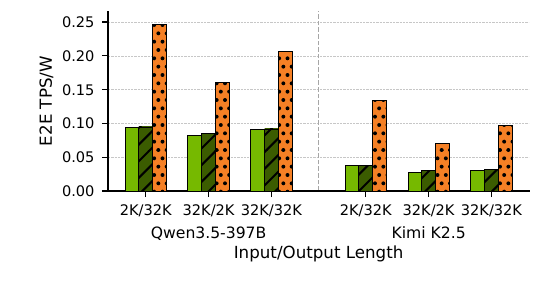}
        \vspace{-1pc}
        \caption{TDP-normalized throughput}
        \label{fig:scaleout_bw_e2e_tps_per_w_b}
    \end{subfigure}

    \begin{center}
        \includegraphics[width=0.9\linewidth]{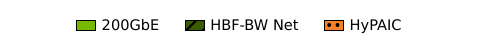}
    \end{center}    
    \vspace{-1pc}
    \caption{\textcolor{black}{Performance comparison based on scale-out interconnect bandwidth improvements.}}
    \vspace{-1pc}
    \label{fig:scale-out}
\end{figure}

\textcolor{black}{As mentioned in Sec.~\ref{subsec:paic_limitations}, running large models using existing PAICs can be achieved by connecting multiple PAICs via a scale-out interconnect, such as 200GbE.
In this scenario, the bandwidth of the scale-out interconnect becomes a performance bottleneck.
Consequently, one might wonder if HyPAIC would become unnecessary if the scale-out interconnect bandwidth were increased to the level of HBF.
However, even if the scale-out interconnect bandwidth bottleneck is resolved, the LPDDR5x bandwidth emerges as the new bottleneck.
Since LPDDR5x bandwidth remains lower than that of HBF, achieving optimal performance remains difficult.
Fig.~\ref{fig:scale-out} demonstrates that the increases in E2E TPS and E2E TPS/W resulting from improved scale-out interconnect bandwidth are negligible.}

\section{Discussion and Future Work}
\label{sec:discussion}

\noindent \textbf{Heat Dissipation: }
Although less severe than in data centers, high power consumption can still be a significant issue for PAICs. 
In particular, the resulting thermal challenges can substantially affect hardware performance. 
Because a PAIC is a small, palm-sized device, physically dissipating heat can be difficult. 
Thus, successfully implementing HyPAIC requires lower HBF power consumption and innovative cooling.

\noindent \textbf{Write Endurance: }
The proposed HyPAIC mitigates HBF’s limited write endurance by storing only immutable model weights in HBF. 
Supporting a broader range of workloads, however, will require substantially improved or actively managed write endurance. 
From a computer architecture perspective, hardware–software co-design can identify workload write patterns and provide mechanisms that allow the device to meet its target lifetime despite runtime writes. 
Alternatively, device-level techniques may improve endurance by trading off density, latency, or retention.

\noindent \textbf{Other NAND Flash Overheads: }
A practical HBF implementation must consider additional overheads such as wear leveling, garbage collection, error correction\textcolor{black}{, read-disturb mitigation, and background data scrubbing. Initial model programming, power cycling, and cold-start page staging may also introduce additional non-steady-state overheads. 
Our architecture-level simulator, an extended version of AIConfigurator~\cite{xu2026aiconfigurator}, does not model these effects because detailed HBF device-management behavior is not yet available. 
Although read-only weight accesses are expected to reduce write-related management activity, these overheads are not assumed to be negligible and remain important future work.}

\section{Related Work}

\noindent \textbf{High Bandwidth Flash: }
Since being unveiled in 2025~\cite{hbf2025}, HBF does not yet exist as a physical device and remains an active area of research.
The scope of research is extensive, ranging from devices to computer architecture.

Device-level HBF research itself primarily focuses on how to implement the device.
Mem-Village proposed a method to implement an HBM and HBF hybrid structure by introducing a glass interposer~\cite{lee2025mem}.
Another study presented a 3D NAND die and cube architecture design optimized for LLM inference to significantly improve memory bandwidth compared to conventional NAND~\cite{park2026architectural}. This study reported that the internal read bandwidth per die exceeds 1 TB/s while maintaining 10$\times$ the capacity of the latest HBM cube.

In the field of computer architecture, research has primarily explored HBF applications~\cite{ha2026h,hsu2026haven} and examined architectural opportunities and challenges during LLM training and inference~\cite{park2026hbm,kyung2026high,wu2026memexplorer,son2026exploring,maeda2026stco,wang2026flashaccel}.
It was demonstrated that system throughput can be improved by applying the HBM-HBF hybrid architecture to cache-augmented generation applications in~\cite{ha2026h} and approximate nearest-neighbor search applications in~\cite{hsu2026haven}.
Architecture exploration studies~\cite{park2026hbm,kyung2026high,wu2026memexplorer,son2026exploring,maeda2026stco,wang2026flashaccel} explored hybrid architecture combinations, HBF access latency, write endurance, etc., to examine the potential advantages and disadvantages of HBF during LLM training and inference.

\noindent \textbf{Low Cost LLM Inference: }
Another line of work reduces the hardware resources needed for large-scale LLM inference. FlexGen~\cite{sheng2023flexgen} demonstrated that models with up to 175B parameters can be served on a commodity GPU using SSD-based offloading. 
InfiniGen~\cite{lee2024infinigen} and ShadowKV~\cite{sun2024shadowkv} used prediction-based prefetching to reduce data-transfer overhead. 

Although prior studies have investigated both HBF and cost-effective LLM inference, to the best of our knowledge, HyPAIC is the first architecture to integrate HBF into a PAIC and run a trillion-parameter model on a single node.

\section{Conclusion}

We propose \textbf{HyPAIC (Hyper-PAIC)}, a novel personal AI computer architecture that integrates HBF and democratizes trillion-parameter LLM inference on a single computer. 
To overcome the inherent memory capacity and bandwidth limitations of existing PAICs, we introduce a heterogeneous memory management method that stores read-only model weights in HBF and mutable data, such as KV caches and activations, in LPDDR5x. 
This placement strategy exploits HBF’s high capacity and bandwidth while mitigating its limited write endurance. 
It is particularly well aligned with the sparse, low batch access patterns of large-scale MoE inference.

Our simulation-based evaluations demonstrate that a single HyPAIC can efficiently execute a 1T-parameter model (Kimi~K2.5), achieving up to \textcolor{black}{1.04}$\times$ higher end-to-end system throughput and \textcolor{black}{3.59}$\times$ better TDP-normalized throughput compared to a baseline configuration requiring \textcolor{black}{8} DGX Sparks.
Furthermore, for small models, the maximum number of concurrent tasks that maintain 50 decode TPS per task increases by up to \textcolor{black}{4}$\times$, significantly enhancing user experience and productivity. 
\textcolor{black}{All these results are achieved using just a single HyPAIC.}
Ultimately, we demonstrate that HBF has the potential to become a promising memory technology in the on-premises AI serving environment.

% \section*{Acknowledgment}
% We used LLMs for limited editing and language refinement.

%%%%%%% -- PAPER CONTENT ENDS -- %%%%%%%%

%%%%%%%%% -- BIB STYLE AND FILE -- %%%%%%%%
\bibliographystyle{IEEEtranS}
\bibliography{refs.bib}
%%%%%%%%%%%%%%%%%%%%%%%%%%%%%%%%%%%%

\appendix
\subsection{AI Use}
The authors used AI models (GPT and Gemini) during manuscript preparation to polish the English writing, assist with Python code for experimental plot generation, and assist with modifications and debugging of the simulator. 
All architectural designs, simulation methodologies, experimental configurations, analyses, and conclusions were developed by the authors. 
The authors reviewed, tested, and verified all AI-assisted code and content and take full responsibility for the correctness of the reported results and the final manuscript.

\end{document}